\documentclass[lettersize,journal]{IEEEtran}
\usepackage{amsmath,amsfonts}
\usepackage{color}
\usepackage{algorithmic}
\usepackage{algorithm}
\usepackage{amssymb}
\usepackage{bm}
\usepackage{array}
\usepackage[caption=false,font=normalsize,labelfont=sf,textfont=sf]{subfig}
\usepackage{textcomp}
\usepackage{stfloats}
\usepackage{url}
\usepackage{verbatim}
\usepackage{graphicx}
\usepackage{cite}
\usepackage[T1]{fontenc}
\usepackage{array}
\usepackage{caption}

\usepackage{geometry}
\newtheorem{theorem}{ \textbf{Theorem}}
\newtheorem{remark}{ \textbf{Remark}}

\newtheorem{game}{ \textbf{Game}}

\begin{document}
\title{Optimal Repair Strategy Against Advanced Persistent Threats Under Time-Varying  Networks}

\author{Zixuan Wang, Jiliang Li, Yuntao Wang, Zhou Su,~\IEEEmembership{Senior Member,~IEEE}, Shui Yu,~\IEEEmembership{Fellow,~IEEE}, and Weizhi~Meng,~\IEEEmembership{Senior Member,~IEEE}
\thanks{This work is supported in part by National Key R\&D Program of China (No. 2022YFB2702800), and in part by NSFC (Nos. 62102305, 62302387, U22A2029, U20A20175), China Postdoctoral Science Foundation under Grant 2023M732820 (Jiliang Li is the corresponding author).\par
Z. Wang, J. Li, Y. Wang, and Z. Su are with the School of Cyber Science and Engineering, Xi'an Jiaotong University, Xi'an 710049, China. \par
  S. Yu is with the School of Computer Science, University of Technology Sydney, Sydney 2007, Australia. \par
  W. Meng is with the Department of Applied Mathematics and Computer Science, Technical University of Denmark (DTU), Copenhagen 2800, Denmark. }
}


\maketitle

\begin{abstract}

Advanced persistent threat (APT) is a kind of stealthy, sophisticated, and long-term cyberattack that has brought severe financial losses and critical infrastructure damages. Existing works mainly focus on APT defense under stable network topologies, while the problem under time-varying dynamic networks (e.g., vehicular networks) remains unexplored, which motivates our work. Besides, the spatiotemporal dynamics in defense resources, complex attackers' lateral movement behaviors, and lack of timely defense make APT defense a challenging issue under time-varying networks. In this paper, we propose a novel game-theoretical APT defense approach to promote real-time and optimal defense strategy-making under both periodic time-varying and general time-varying environments. Specifically, we first model the interactions between attackers and defenders in an APT process as a dynamic APT repair game, and then formulate the APT damage minimization problem as the precise prevention and control (PPAC) problem. To 
derive the optimal defense strategy under both latency and defense resource constraints, we further devise an online optimal control-based mechanism integrated with two backtracking-forward algorithms to fastly derive the near-optimal solution of the PPAC problem in real time. Extensive experiments are carried out, and the results demonstrate that our proposed scheme can efficiently obtain optimal defense strategy in 54481 ms under seven attack-defense interactions with 9.64$\%$ resource occupancy in stimulated periodic time-varying and general time-varying networks.
Besides, even under static networks, our proposed scheme still outperforms existing representative APT defense approaches in terms of service stability and defense resource utilization.
\end{abstract}

\begin{IEEEkeywords}
Cybersecurity, advanced persistent threat (APT), node-level epidemic model, APT repair game, time-varying networks. 
\end{IEEEkeywords}

\section{Introduction}
\IEEEPARstart{R}{ecently}, many well-protected organizations, such as multinational companies or government departments, have been subjected to various cyber attacks that cause huge financial losses, disclosure of business secrets, and even damages to nation-critical infrastructures \cite{WorldwideInfrastructureSecurity,wang2022survey,wang2023survey}. According to 360 Security's statistics \cite{360AnQuanDaNao}, advanced persistent threats (APT) such as Stuxnet, Duqu, Flame, and Gauss, account for nearly $60\%$ of the attacks on government departments and multinational corporations in the past two years. 

To launch a typical APT attack, a team of collusive hackers generally use sophisticated, distributed, and stealth attack technologies to (i) gain illicit access to the target system (e.g., via malware), (ii) infiltrate target organization via privilege escalation and lateral movement attacks, and then (iii) achieve illegal purposes.
Essentially, APT attacks are characterized by long-term attack cycles, highly sophisticated changes, and unpredictable lateral movements, thereby causing highly complex attack patterns and difficulties in attack path tracing and system reparation. 

\par
 Due to the characteristics and hazards of APT attacks, how to defend against APT attacks has become a hot topic of research in academia and industry. There already have been various works \cite{yang2018risk,yangEffectiveQuarantineRecovery2021,yangEffectiveRepairStrategy2019,li2018intelligence,li2018defending,zhang2019mathtt,feng2019dynamic} focusing on APT defense, and many of them are based on the game theory. 
 For example, Yang \emph{et al.} \cite{yang2018risk} proposed a game-theoretical APT risk management strategy for defenders (e.g., the security department of an organization) to optimally deploy defense resources with mitigated APT attack loss, in which both the assets of the organization and the behavioral returns of the adversary are quantified. 
Li \emph{et al.} \cite{li2018intelligence} presented a Lyapunov-based security-aware defense mechanism and employed the game theory to seek the equilibrium between the network defenders and APT attackers for optimally allocating the defensive resources with maximized system utility. Yang \emph{et al.} \cite{yangEffectiveRepairStrategy2019} modeled the APT repair problem as an APT defense game and presented a greedy algorithm to seek the Nash equilibrium as the optimal defensive response strategy.\par
However, existing works mainly assume that the network topology of an organization (e.g., government department and operator of an IoT network) remains stable during APT defense, which is inapplicable to general scenarios with dynamic network topology (e.g., vehicular network \cite{wang2022task,wang2021blockchain} and UAV network \cite{wang2023seal,wang2023secure}). Many researchers already have recognized the impact of network typology's dynamicity on real-world attacks and defenses, especially in the field of Internet of Things (IoT) \cite{xiao2018security,nguyen2021deep,wang2020bc}. Besides, the highly dynamic network topology can result in more complex attackers' lateral movement patterns, as well as a necessity for online low-latency APT defense strategy-making under the constraint of defensive resources. For example, the dynamic nature of satellite networks, exemplified by the successful Hack-A-Sat competition \cite{Hack-A-Sat} on Aug 18, 2023 at DEF CON \cite{DEFCON}, highlights how the constantly changing topology can be exploited by attackers, compromising even well-defended satellites (details refer to Appendix A). Particularly, the following key challenges must be addressed to practically deploy the game-theoretical APT countermeasures.

\par

\begin{figure*}[!t]
\centering
\subfloat[]{\includegraphics[width=2.3in]{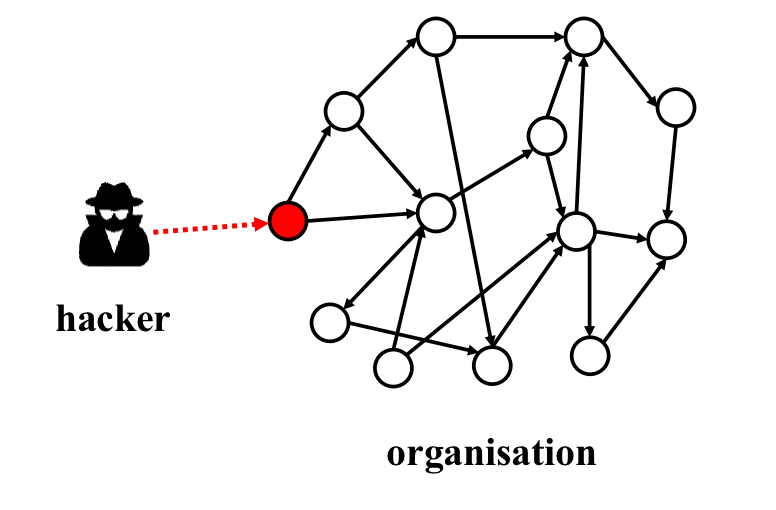}%
\label{fig_first_case}}
\hfil
\subfloat[]{\includegraphics[width=2.2in]{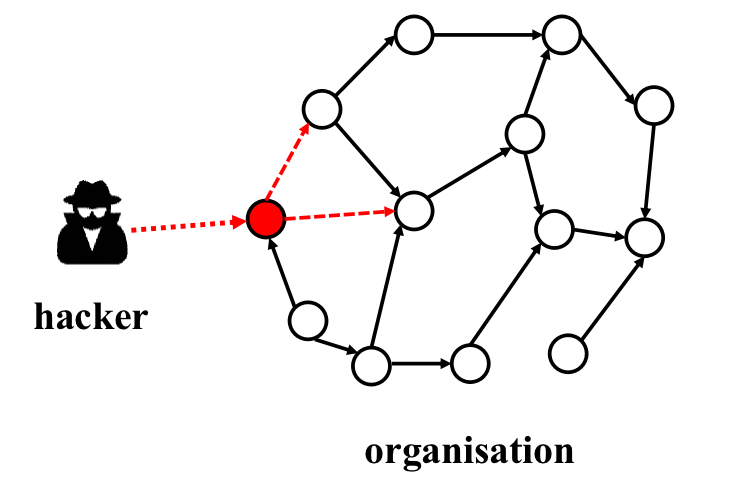}%
\label{fig_second_case}}
\hfil
\subfloat[]{\includegraphics[width=2.2in]{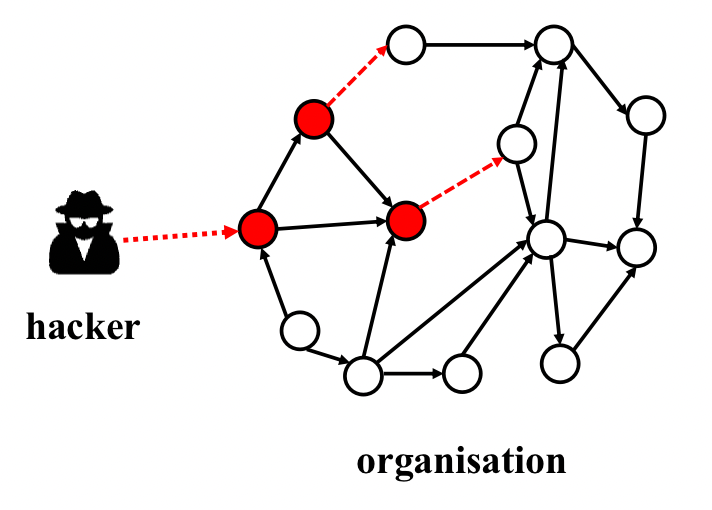}%
\label{fig_third_case}}
\hfil
\caption{ The lateral movement attack on the entire dynamic network organization after the attacker obtains the ingress machine. The red nodes indicate the machines that have been hijacked. The red dashed lines indicate the potential targets of the lateral movement attack. The change of topology in (a)-(c) indicates the change of network topology.}
\label{movement}
\end{figure*}

\begin{itemize}
      \item \textit{Dynamical defense resources}. 
      Typically, to recover from the APT attack and offer normal services, an organization's defensive resources (e.g., audit and bandwidth resources) are usually limited. Moreover, depending on practical applications, the organization's defensive resources might be dynamic in terms of space and time. 
      
      \item \textit{Inefficiency in time-varying networks}. As shown in Fig.~\ref{movement}, the lateral movement attack (i.e., the attacker penetrates other organization nodes through its compromised nodes) under a dynamic network can severely damage the network's functionality. Furthermore, the time-varying network topology complicates lateral movement attacks, resulting in higher uncertainty and inefficiency in defensive resource allocation. 
      
      \item \textit{Lack timely defense}. Since the network topology changes over time, the defensive behaviors of defenders should be timely adapted and enforced on the entire network. Besides, considering the impact of defenders' behaviors in the time-varying network topology, there exists a curse of dimensionality due to the huge space of state and action sets.   
    
  \end{itemize}
 
To bridge the research gap, this paper proposes a fast and efficient APT defense strategy tailored to time-varying networks under both defense resource limitations and network dynamics.
Specifically, we model the interactions between APT attackers and defenders (i.e., the network operator) under highly dynamic environments as a sequential APT repair game with the aim of minimizing the damage of the APT attack to the organization. We regard the APT attack process as a finite time horizon divided into multiple time slots, as shown in Fig. \ref{movement}. The interactions between APT attackers and defenders in each time slot are modeled as a subgame of the dynamic APT repair game. Based on the sequential APT repair game, an optimization problem, i.e., precise prevention and control (PPAC), is formulated under both latency and defense resource constraints.
 
 Considering the time-varying network topology, threat impacts, hosts' differentiated states, and spatio-temporal variability of defensive resources, we devise an online fast searching mechanism based on optimal control to derive the near-optimal solution of the PPAC problem in real time. We summarize the main contributions as follows.

\begin{itemize}
      \item \textit{Dynamic APT Repair Problem Modeling}. We model the interactions between APT attackers and defenders in each time slot as a subgame of the dynamic APT repair game. The sequential APT repair game is formulated as a precise prevention and control (PPAC) optimization problem, whose solution is the online optimal defense strategy. To the best of our knowledge, this is the first attempt at the defense strategy against APT attacks under time-varying networks.
      \item \textit{Efficient and Practical Solving Method}. To solve the PPAC problem, we present a practical online fast strategy searching mechanism to seek the optimal APT defense strategy which minimizes the impact of APT. It consists of two backtracking-forward algorithms and an optimal system. The two backtracking-forward algorithms, i.e., the threat rate grading (TRG) algorithm and the recovery rate grading (RRG) algorithm are developed to compute the nodes' states at each moment. Based on nodes' states, the designed optimal system can overcome the challenges caused by topology changes and complex APT attack patterns to derive the optimal defense strategy.
   
      \item \textit{Extensive Experimental Validation}. Experiments on three scenarios (i.e., static, periodic time-varying, and general time-varying networks) demonstrate the feasibility and effectiveness of the proposed scheme. Results from experiments validate that the proposed scheme can reduce the impact of APT attacks on system utility and save defense resources in time-varying dynamic networks. Compared with existing representatives, our proposed approach can effectively enhance service stability to 100$\%$ and defense resource utilization to 75.50$\%$ in stimulated time-varying networks, thereby greatly enhancing both the service utility and defense performance.
      
  \end{itemize} \par

 The rest of the paper is organized as follows. Section II reviews the related works, and Section III introduces the system model. Section IV introduces the sequential APT repair game model and formulates the PPAC problem. Section V introduces the online fast strategy searching mechanism to solve the PPAC problem, and Section VI evaluates the performance of the proposed scheme. Section VII summarizes this work using conclusions.  

\section{Related Work}
This section presents a comprehensive review of attack propagation in APT scenarios and provides an in-depth analysis of the existing literature on defense strategies. We review the attack dissemination related literature (in Sect.~II.A), game-theoretical APT defense (in Sect.~II.B), and optimal control-based APT defense (in Sect.~II.C).

\par

\subsection{Attack Dissemination}
The attack dissemination, also referred to as the lateral movement phase \cite{AttackMatrix} of a cyber attack, is intricately influenced by the network topology. Extensive research has explored the complex dynamics of propagation, particularly within the domains of synchronization and consensus in static or dynamic multi-agent systems \cite{lu2005time}.
\par
Kephart \emph{et al.} \cite{kephart1992directed} and Staniford \emph{et al.} \cite{staniford2002own} are among the first to identify that the propagation of Internet worms exhibits similarities to the spread of epidemics, posing a significant threat to system security. Building upon their pioneering work, subsequent research in the field has continued to explore the analogy between cyber attacks and epidemic spread. For instance, Wang \emph{et al.} \cite{wang2009understanding} apply epidemic models to study the propagation of viruses in mobile phones through multimedia messaging services or Bluetooth connections. Cheng \emph{et al.} \cite{cheng2010modeling} extend the application of epidemic models by simulating the propagation of malicious software within generalized social networks comprising both non-local and local connections. Their research further validates the suitability of epidemic models in capturing the dynamics of attack dissemination in such networks. 
\par
Another notable contribution comes from Darling \emph{et al.} \cite{darling2008differential}, who demonstrate the efficacy of ordinary differential equations (ODEs) in approximating the dynamics of information dissemination from Markov chains. And, Lapidus \emph{et al.} \cite{lapidus1971numerical} prove that the backtracing-forward algorithm demonstrates exceptional performance in solving such problems. Therefore, the utilization of Markov chains in conjunction with node states to construct epidemic models \cite{zino2021analysis} has extensive applications in APT defense \cite{yangEffectiveRepairStrategy2019,yangEffectiveQuarantineRecovery2021}.
\subsection{Game-theoretical APT Defense}
Recently, various research efforts have focused on using game theory to find efficient defense strategies for APT attacks. The pioneering work of using game theory (the FlipIt game) to model the payoffs and perform equilibrium analysis for both attackers and defenders was proposed by van Dijk \emph{et al}. \cite{van2013flipit}. 

Subsequently, the FlipIt game is used in various cyber situations \cite{zhang2019mathtt,feng2019dynamic}, which contains multiple resources, information leakage, etc. In these applications, \cite{huang2019adaptive} uses the PBNE to discuss the APT-related defense strategies. 

Zhu \emph{et al}. \cite{zhu2018multi} categorize APT attacks into three distinct phases and introduce a comprehensive framework based on matrix and sequential games. By iteratively determining the optimal defense strategy at each stage, the overall best defense strategy is derived.
\par
In addition to applying game theory to study some general aspects of APT attacks, several significant works have been conducted on resource-limited defense strategies. These strategies focus on optimizing resource utilization and developing effective countermeasures to combat APT threats. This kind of strategies is commonly known as the repair strategy \cite{yangEffectiveRepairStrategy2019}. To identify the most optimal approach, the concept of a repair game has been proposed. In line with this objective, Ye \emph{et al}. \cite{yeDifferentiallyPrivateGame2021} study a new differential game based on network spoofing and developed differential privacy methods for Nash equilibrium. But they ignore the lateral movement of each compromised host. 
Yang \emph{et al}. \cite{yangEffectiveRepairStrategy2019} discuss the control strategy approach for lateral movement in traditional network architectures with a differential game approach. They give a potential Nash equilibrium under this game from the perspective of the defender and give a reference for the defender's defense strategy. However, they only briefly discuss repair methods while ignoring the diversity of repair methods. In \cite{yangEffectiveQuarantineRecovery2021}, Yang \emph{et al}. propose the idea of isolation zones to solve the APT repair problem in a more reasonable way which constructs a two-order dynamical system to solve the control set for the QAR (quarantine and recovery) problem. \par
However, all the works mentioned above neglect the dynamic nature of topology in dynamic networks, thereby, cannot be applied to dynamic networks.

\subsection{Optimal Control Model for APT Defense} 
Besides the game-theoretical approach, various works use optimal control to defend against APT attacks. Optimal control theory is used to find the required control law to make a given system perform optimally and is widely used in cybersecurity. The defender simply has to be aware of the average values of a few factors connected to the APT when modeling the APT defense problem using optimal control. As compared to other methods, estimating these factors requires less real APT data, which the APT attack and defense exercises \cite{diogenes2018cybersecurity} can be used to estimate these factors. Li \emph{et al}. proposed \cite{li2018defending} the optimal control methods to solve the APT problem. They used a node-level contagion model to evaluate changes in the state of nodes in the maintenance scheme, which modeled the APT response problem as an optimal control problem. In \cite{zhao2019minimizing}, Zhao \emph{et al.} modeled the isolation-conversion problem as an optimal control problem to find the optimal isolation strategy to reduce the negative impact of rumors. Inspired by \cite{zhao2019minimizing}, Yang \emph{et al}. \cite{yangEffectiveQuarantineRecovery2021} further considered the isolation operations essential to the APT response process. \par
However, existing works on APT defense mainly focus on the networks with stable structures (e.g., the internal network inside an organization) and are inapplicable to time-varying networks (e.g., vehicular networks). Besides, these works ignore the impacts of defense behaviors on the utility of the network operator, which eventually deteriorates the defensive effectiveness. Different from existing works, our work jointly considers the dynamic topological characteristics of the nodes and the impacts on the service utility in time-varying networks to develop fast and effective APT defensive strategies.

\section{System Model}
This section introduces the system model, including the network model, lateral movement attack model, and potential APT impact model. The key notations in this paper are listed in Table~\ref{tab:Notion}.

\subsection{Network Model}\label{dyn APT model}
In the APT attack-defense scenario, it involves an attacker (denoted by $\mathbb{A}$) and a network operator (i.e., defender, denoted by $\mathbb{D}$) who operates \emph{V} number of hosts. Here, $\mathbb{A}$ can be an attacker group conspired to launch the APT attack, and $\mathbb{D}$ can be the cyber security department of an organization (i.e., network operator). At the same time, \emph{V} also represents the organization's valuable assets (e.g., business secrets). Since the defender is far-sighted\footnote{As the APT attack is persistent, the defender aims to reduce the total loss in the whole APT attack process.} and aims to minimize the loss during the APT process, we study the variation of APT lateral movement attack and the expected impact of APT in a finite time horizon, which is divided into multiple time slots. The time horizon of the defense process is denoted as $[0, T]$, which is determined by the length of the APT process. We denote the sequence of the APT process as 
\begin{equation}
\label{eq_T}
\bm{\hat{T}}=\{A_1=(0,t_1),A_2=(t_1,t_2),...,A_m=(t_{m-1},t_{m}\},
\end{equation}
where $t_j (1 \leq j \leq m)$ means the periods of the APT attack. Due to the characteristics of the time-varying network, we represent the periods of APT by the topology adjacency matrix of the network. Each element $A_j (1 \leq j \leq m)$ is a topology adjacency matrix. For example, if $0\leq t \leq t_1$, the adjacency matrix $A(t)=A_1$. 
\begin{table}\scriptsize
    \centering
    \caption{Key Notations 
    }
    \resizebox{\linewidth}{!}{
    \begin{tabular}{m{2cm}<{\centering}|m{5cm}<{\centering}}
    \hline
      Notation  &  Description     \\ \hline
      $G(t)=\{V,E(t),A(t)\}$ & The topological graph of the server network, $\xi$ denotes the edge set of the graph at time t, $A(t)=(a_{ij}^t)_{N \times}$ denotes the adjacency matrix of the $G$ at time t.    \\ \hline
      $\bm{\hat{T}}$ & The topology schedule of the task. $\bm{\hat{T}}=\{A_1=(0,t_1),A_2=(t_1,t_2),...,A_m=(t_{m-1},t_{m}\}$   \\ \hline
      T & Time to perform the task.    \\ \hline
      $X_i(t)=\{0,1,2,3\}$ &The node $v_i$ is healthy, mild patients, severe patients, and quarantined at time t, respectively. The vector $\mathbf{X}(t)=(X_1(t),...,X_N(t))$ represented the state of network at time t.\\ \hline
      $\beta_i(t)$ & The average rate of direct infiltration that compromises a healthy node $v_i$. Let $\bm{\beta}(t)=(\beta_1(t),...,\beta_N(t))$\\ \hline
      $\alpha_i(t)$ & The average rate of lateral movement from any nearby compromised APT node that compromises a healthy node $v_i$.  Let $\bm{\alpha}(t)=(\alpha_1(t),...,\alpha_N(t))$.  \\ \hline
      $\lambda_i(t)$ & The average rate at which compromised node $v_i$ is classified as mild or severe. Let $\bm{\Lambda}(t)=(\lambda_1(t),...,\lambda_N(t))$. The function $\mathbf{\Lambda}$ represents the threaten scheme.    \\ \hline
      $\delta_i(t)$ & The average rate at which compromised node $v_i$ is quarantined or not. Let vector $\bm{\Delta}(t)=(\delta_1(t),...,\delta_N(t))$. The function $\mathbf{\Delta}$ represents the quarantined scheme.    \\ \hline
      $\gamma_i(t)$ & The average rate at which severe patients node $v_i$ is recover to mild patients or healthy. Let vector $\bm{\Gamma}(t)=(\gamma_1(t),...,\gamma_N(t))$. The function $\mathbf{\Gamma}$ represents the recovery scheme.   \\ \hline
      
      $a_i^1, a_i^2$ & The average consumption per unit time of $M_i$ and $S_i$. Vectors $\bm{a^1}=(a^1_1,...,a^1_N)$ and $\bm{a^2}=(a^2_1,...,a^2_N)$ denote resource impact value vector of the network, respectively. \\ \hline

      $b_i$ & The average impact of services impact per unit of time. is referred to as a precise prevention and control scheme. $\bm{b}=(b_1,...,b_N)$ denotes services impact value vector of the network.\\ \hline

      $\phi_i$ & The quarantine cost function of the node $v_i$. Vector $\bm{\phi}=(\phi_1,...,\phi_N)$ represents the quarantine cost function of the network. \\ \hline

      $\varrho^1_i,\varrho^2_i$ & The recovery cost function to $M_i$ and $S_i$. Vector $\bm{\varrho^1}=(\varrho^1_1,...,\varrho^1_N)$ and vector $\bm{\varrho^2}=(\varrho^2_1,...,\varrho^2_N)$ represent the recovery cost of the network. \\ \hline

     $R, T$ & The output of algorithm TRG and RRG, which represent the threaten states and recovery states of the network. \\ \hline

     $\bm{U_n},\bm{U}$ & $\bm{U}$ represents the utility of the entire network. $\bm{U_n}$ represents the requirement of maintaining minimum service effectiveness. \\ \hline

     $\bm{E}_0$ & $\bm{E}_0$ represents the average rate of the initial state of nodes.  \\ \hline

     $\bm{w}$ & $\bm{w}(t)=(\mathbf{\Lambda}(t),\mathbf{\Delta}(t),\mathbf{\Gamma}(t))$ is referred to as a precise prevention and control scheme.    \\ \hline 
    \end{tabular}
    }
     \label{tab:Notion}
\end{table}
 We denote the links of network's devices as $G(t)=\{V,E(t),A(t)\}$, where $V=\{v_1,v_2,...,v_N\}$ stands for the sets of all devices and the edge $ \{v_i, v_j\} \in E(t)$ stands for the relation between devices $v_i$ and $v_j$ at time $t$. Let $A(t)=(a_{ij})_{N \times N}$ denote the adjacency matrix of network $G$ at time $t$, $a_{ij}=1$ or $0$ represents that there exist a link between $v_i$ and $v_j$ or not, respectively.\par

 Assume the defender discovers the APT warning at time slot $t=0$ and replies within the allotted time window $[0, t]$ by carrying out several defense activities. Each node has four possible states during the response period at any given time slot: $healthy$, $mild$, $severe$, and $quarantined$. The $quarantined$ state stands for the nodes that break all the links. The $healthy$ state represents the nodes that do not suffer from attacks. The $mild$ state stands for the compromised nodes whose impact on the network is acceptable. In other words, the $mild$ nodes bring more revenue than the loss they caused. The $severe$ state represents the compromised nodes whose impact on the network is fatal, which means they need to be repaired immediately. For each node $v_i$, let $X_i(t)=0,1,2$, and $3$ denote that $v_i$ is $healthy$, $mild$, $severe$, and $quarantined$ at time $t$, respectively. Then the state of each node at time $t$ can be denoted as:
\begin{equation}
\label{eq_X}
\mathbf{X}(t)=(X_1(t),X_2(t),...,X_N(t)).
\end{equation}
We provide the following set of notations to express the network's evolution process for the $\mathbf{X}(t)$ variable.
 \begin{enumerate}
     \item We define the PPAC control factor as a $3N$-dimensional function $\bm{w}$, i.e.,
    \begin{equation}
    \label{eq_w}
    \bm{w}(t)=(\mathbf{\Lambda}(t),\mathbf{\Delta}(t),\mathbf{\Gamma}(t)). \quad 0\leq t\leq T
    \end{equation}
     $\bm{W}=(\bm{w}(1),...,\bm{w}(j))$ represents the control vector of whole APT process. Here, the average rate is a fundamental concept in stochastic process theory \cite{stewart2021introduction} whose reciprocal represents the mean time before the event occurs.
      \item The average rate at which each compromised node $v_i$ is categorized as mild or severe is represented by the notation $\lambda i(t)$ ($\lambda_i(t)$ is severe, $1-\lambda_i(t)$ is mild). We refer to the $N$-dimensional function $\mathbf{\Lambda}$ defined by
\begin{equation}
\label{eq_lambda}
\bm{\Lambda}(t)=(\lambda_1(t),...,\lambda_N(t)), \quad 0\leq t\leq T
\end{equation}
as the threat vector.
      \item The average rate of quarantining for each severe node $v_i$ at time $t$ is represented by the notation $\delta_i(t)$. We refer to the $N$-dimensional function $\mathbf{\Delta}$ defined by
\begin{equation}
\label{eq_delta}
\bm{\Delta}(t)=(\delta_1(t),...,\delta_N(t)), \quad 0\leq t\leq T
\end{equation}
as the quarantined vector.
\item The average rate at which a confined node $v_i$ becomes healthy or mild at time $t$ is expressed as $\gamma_i(t)$ for each quarantined node $v_i$. We refer to the $N$-dimensional function $\mathbf{\Gamma}$ defined by
\begin{equation}
\label{eq_gamma}
\bm{\Gamma}(t)=(\gamma_1(t),...,\gamma_N(t)), \quad 0\leq t\leq T
\end{equation}
as the recovery vector.
     \item The healthy nodes of the network that suffered from attacks are defined as follows. The healthy node $v_i$ becomes compromised as a result of APT penetration, as represented by $\beta_i(t)$. The probability that the healthy node $v_i$ becomes compromised due to lateral movement from surrounding compromised nodes is represented by the average infection rate $\alpha_i(t)$. Two statistical properties of the task are $\beta_i(t)$ and $\alpha_i(t)$, which are denoted as $\bm{\beta}(t)=(\beta_1(t),...,\beta_N(t))$ and $\bm{\alpha}(t)=(\alpha_1(t),...,\alpha_N(t))$, respectively. Note that $\bm{\beta}(t)$ and $\bm{\alpha}(t)$ can be calculated using APT attack-defense maneuvers \cite{diogenes2018cybersecurity}. 
  \end{enumerate} 
  \par
  Based on the above notations, we can obtain the state transition of the node $v_i$ at time slot $t$ as shown in Fig. \ref{fig_2} through Markov chain theory. The details of the state transition model are shown in Section \ref{lateral mov model}.
\subsection{Modelling the Lateral Movement Attack}\label{lateral mov model}
The time-varying network topology makes lateral movement attacks more complex and difficult to predict. To accurately portray the impact of the lateral attack, we adapt the Markov process and causal analysis to obtain a response model.\par
Based on the \cite{zino2021analysis}, we can get the state transition function:
\begin{equation}
\label{eq_transe}
\Dot{x}(t)=(1-x_i)\beta_i(t)\sum_{j=1}^na_{ij}(t)x_j-\delta_i(t)x_i(t).
\end{equation}
\begin{figure}[!h]
\centering
\includegraphics[width=2.5in]{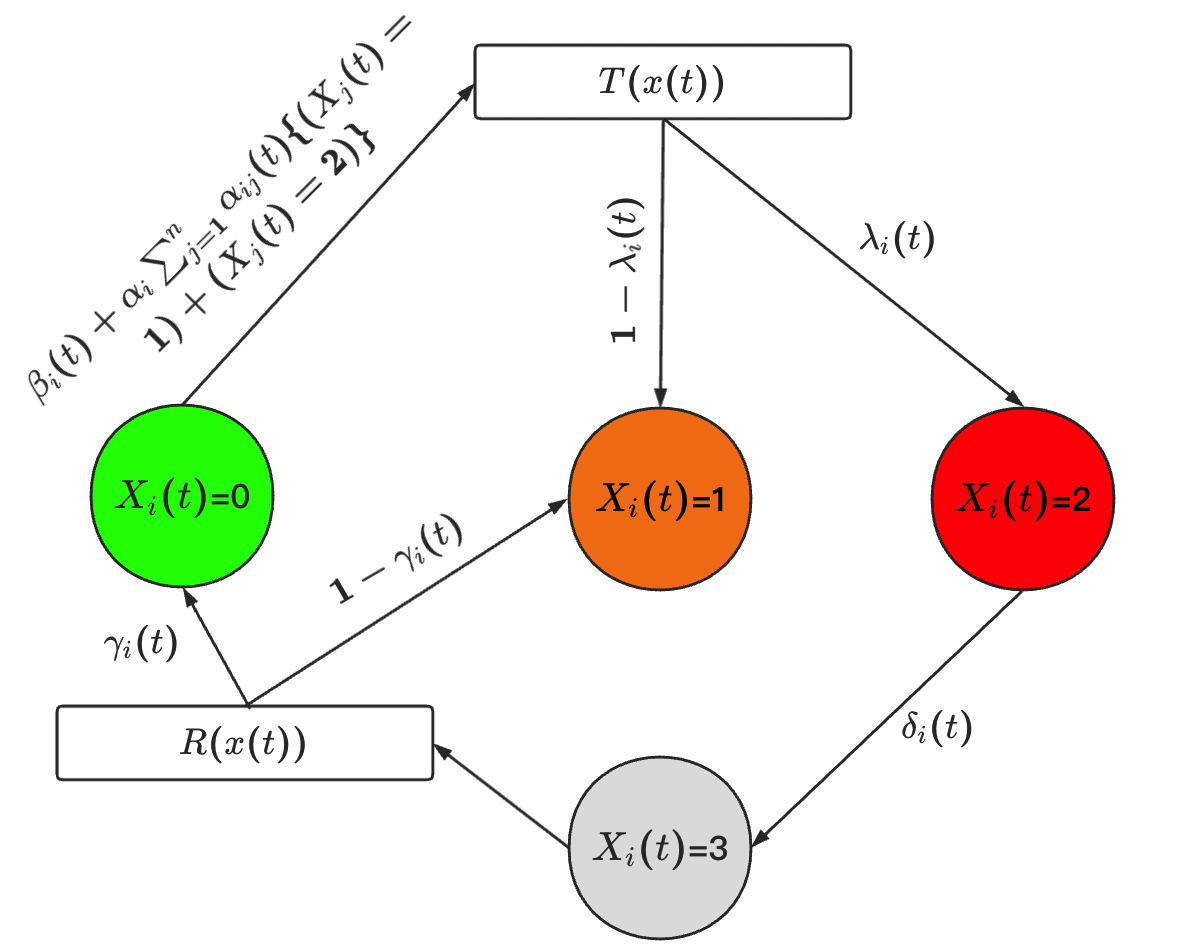}
\caption{The state transition of node $v_i$ at time $t$.}
\label{fig_2}
\end{figure}

Considering both direct infiltration attacks and lateral movement attacks, we define $\bm{X}_i(t)$ as:
\begin{equation}
\label{eq_trans}
\bm{X}_i(t)=\beta_i(t)+\alpha_i\sum_{j=1}^na_{ij}(t)\{(X_j(t)=1)+(X_j(t)=2)\}.
\end{equation}
The state transition process is shown at time $t$ in Fig.~\ref{fig_2}. We use $H_i(t)$, $M_i(t)$, and $S_i(t)$ to denote the probabilities of node $v_i$ being $healthy$ state, $mild$ state, and $severe$ state at time $t$, respectively. Each state can be expressed as:
\begin{equation}
\label{eq_problity}
\begin{split}
H_i(t)=Pr\{ X_i(t)&=0\},\quad M_i(t)=Pr\{ X_i(t)=1\}, \\
 S_i(t)&=Pr\{ X_i(t)=2\}.    
\end{split}
\end{equation}
 If the node $v_i$ is in $quarantined$ state at time $t$, the probability is $1-H_i(t)-M_i(t)-S_i(t)$. 
Vector $\bm{E(t)}$ represents the expected state of the network at time $t$. In practice, we can get the initial state $\bm{E}(0)$ by the APT detection program in \cite{xiong2022conan,ghafir2018detection,milajerdi2019holmes}. The vector $\bm{E(t)}$ can be expressed as:
\begin{small}
   \begin{equation}
\label{eq_pvector}
\begin{split}
\bm{E(t)}=(&H_1(t),...,H_N(t),M_1(t),...,M_N(t),S_1(t),...,S_N(t)).
\end{split}
\end{equation} 
\end{small}\par
 Therefore, we have the following theorem to describe the network's expected state evolution.
\begin{theorem}\label{themo_1}
The network's expected state evolves according to the following differential dynamical system:
\begin{equation}\label{eq_differential dyna sys}
\begin{aligned}
 \begin{cases}
     \frac{dH_i(t)}{dt}&=\gamma_i(t)[1-M_i(t)-S_i(t)-H_i(t)]- H_i(t)* \\
     & [\beta_i(t)+\alpha_i\sum_{j=1}^na_{ij}(t)M_j(t)\\
     &+\alpha_i\sum_{j=1}^na_{ij}(t)S_j(t)]  \\
     \frac{dM_i(t)}{dt}&=H_i(t)[\beta_i(t)+\alpha_i\sum_{j=1}^na_{ij}(t)M_j(t)\\
     &+\alpha_i\sum_{j=1}^na_{ij}(t)S_j(t)]-\lambda_i(t)M_i(t) \\
     \frac{dS_i(t)}{dt}&=\lambda_i(t)M_i(t)-\delta_i(t)S_i(t)\\
     \bm{E}(0)&=\bm{E}_0
 \end{cases}
   \end{aligned}
\end{equation}
\end{theorem}
\begin{IEEEproof}
We use $\mathbb{E}$ to denote the mathematical expectation. The expected rate at which $healthy$ state node $v_i$ transforms to $severe$ state or $mild$ state at time $t$ is
\begin{small}
\begin{equation}
\begin{aligned}
    \mathbb{E}\biggl( \bm{X}_i(t)&=\beta_i(t)+\alpha_i\sum_{j=1}^na_{ij}(t)\{(X_j(t)=1)+(X_j(t)=2)\} \biggr)\\
    &= \bm{X}_i(t)=\beta_i(t)+\alpha_i\sum_{j=1}^na_{ij}(t)(M_i(t)+S_i(t)),
\end{aligned}
\end{equation}
\end{small} 
\\and the derivation of the conversion process of other states in Fig.~\ref{fig_2} is the same. Hence, we can get the Eq. (\ref{eq_differential dyna sys}).
\end{IEEEproof}
\begin{remark}
Theorem \ref{themo_1} illustrates the differential equations of each state we defined by finding the expectation for the state function Eq. (\ref{eq_trans}) with different condition restrictions. For example, $healthy$ state nodes are composed of attack and recovery two parts at time slot $t$. On the attack part, each $healthy$ node compromise to $mild$ or $severe$ state by direct attack and lateral movement attack, and the compromised probability is $\beta_i(t)+\alpha_i\sum_{j=1}^na_{ij}(t)(M_i(t)+S_i(t))$. On the recovery part, each $quarantined$ node transforms to the $healthy$ state by the defenders' recovery operations whose recovery rate is $\gamma_i(t)$. 
\end{remark}
\subsection{Modelling the Expected Impact of APT Attack}\label{impact}
For the stealthiness of attacks, the attacker hacks the organization slowly via insecure hosts. Therefore, the amount of loss per unit of time of the organization is limited. In this paper, we assume that the attacker will intelligently exploit different insecure hosts to expand the erosion of the organization.\par
We consider the following three main factors of organizational loss caused by the APT attack (i.e., the APT impact on the organization). 1) The consumption of resources by the hijacked hosts. 2) The service utility loss of the attacked host and its impact on the overall service. 3) The cost of adapting the defense strategies to control APT attacks (e.g., cutting off the connection between hosts or system updates).\par
\textbf{1) Resource consumption}. For each compromised node $v_i$, we define $a^1_i$ and $a^2_i$ as the average consumption per unit time of $M_i$ and $S_i$, respectively. Let $\bm{a^1}=(a^1_1,...,a^1_N)$ and $\bm{a^2}=(a^2_1,...,a^2_N)$ denote resource impact value vector of the network, respectively. In practice, the average consumption can be estimated by the resource per unit of time (e.g., bandwidth usage per unit of time and WAF placement per unit of time). Therefore, the following theorem is introduced to define resource consumption in a time period.
\begin{theorem}\label{theo2}
Due to the consumption of resources, the expected consumption in time horizon $[0, T]$ is
\begin{equation}\label{eq_lost source}
    \begin{aligned}
   L(\bm{w}) =\int_0^T\sum_{i=1}^N[&a^1_i M_i(t,\bm{w}(t),G(t)) \\ &+a^2_i S_i(t,\bm{w}(t),G(t))]dt.
   \end{aligned}
\end{equation}
\begin{IEEEproof}
The average consumption in the infinitesimal time horizon $[t, d+dt)$ is $a^1_i$ ($Xi(t) = 1$) and $a^2_i$ ($Xi(t) = 0$). The expected consumption in $[t, d + dt)$ is $a^1_i M_i(t,\bm{w}(t),G(t))dt$ + $a^2_i S_i(t,\bm{w}(t),G(t))dt$. 
\end{IEEEproof}
\end{theorem}
\begin{remark}
 From the consumption history data of the specified hosts after being attacked, defenders can obtain the $M_i$ and $S_i$ state hosts' average consumption $a^1_i$ and $a^2_i$, respectively. In this paper, we assume the average consumption is constant in the infinitesimal time. Through the parameters $a^1_i$ and $a^2_i$, we can define the consumption in the infinitesimal time of one node as $a^1_i M_i(t,\bm{w}(t),G(t))+a^2_i S_i(t,\bm{w}(t),G(t))$. In this way, the integral formulation can calculate resource consumption for a given time period.
\end{remark}
\par
\textbf{2) Service utility loss}. For each quarantined node $v_i$, we define $b_i$ as the average impact of service per unit of time. $\bm{b}=(b_1,...,b_N)$ denotes services impact value vector of the network. In actuality, the vector can be calculated by examining the server's traffic flow. Based on the above, Theorem \ref{theo3} is introduced to define service utility loss in a time period.
\begin{theorem}\label{theo3}
Due to the impact of services, the expected impact in the time horizon $[0, T]$ is
\begin{small}
\begin{equation}\label{eq_effect function}
\begin{aligned}
   E(\bm{w}) =\int_0^T&\sum_{i=1}^N [b_i(1-M_i(t,\bm{w}(t),G(t))\\
   -&S_i(t,\bm{w}(t),G(t))-H_i(t,\bm{w}(t),G(t)))]dt.
\end{aligned}
\end{equation}
\end{small}
\begin{IEEEproof}
The proof of service's impact is comparable to the proof in Theorem \ref{theo2}, and we omit it here.
\end{IEEEproof}
\end{theorem}
\begin{remark}
In our state evolution model, we perform quarantine operations based on the threat rate for the hijacked hosts. So the service utility loss for a given period is only considered for the nodes in the quarantine state, and it is simple to get the service parameter $b_i$ of each host for the defender. Through the parameters $b_i$, we define the service utility loss in the infinitesimal time of one node as $(1-M_i(t,\bm{w}(t),G(t))-S_i(t,\bm{w}(t),G(t))-H_i(t,\bm{w}(t),G(t)))$. The integral formulation can calculate the service utility loss for a given period.
\end{remark}
\par
\textbf{3) Expected repair cost}. For each $severe$ state node, $\phi_i(\delta)$ denotes the cost of quarantining at the average rate $\delta$. We define $\phi_i$ as the quarantine cost function of the node $v_i$. Therefore, the quarantine cost function of the network is represented as a vector $\bm{\phi}=(\phi_1,...,\phi_N)$. The function $\phi_i$ is strictly increasing and $\phi_i(0) = 0$. In practice, we can approximate $\bm{\phi}$ through statistical regression of APT attack-defense maneuvers \cite{diogenes2018cybersecurity}. The dynamic quarantine schedule, denoted by
\begin{equation}\label{eq_quarantine scheme}
\begin{aligned}
   (\phi_1(\delta_1(t)),...,\phi_i(\delta_N(t))) \quad 0\leq t \leq N,
\end{aligned}
\end{equation}
is decided directly by the defender. \par
For each quarantined node $v_i$, we define $\varrho^1_i,\varrho^2_i$ as cost functions to $M_i$ and $S_i$, respectively. Therefore, the recovery cost of the network is represented as vector $\bm{\varrho^1}=(\varrho^1_1,...,\varrho^1_N)$ and vector $\bm{\varrho^2}=(\varrho^2_1,...,\varrho^2_N)$. The dynamic recovery schedule is defined as
\begin{equation}\label{eq_recovery scheme}
\begin{aligned}
   &(\varrho^1(\gamma_1(t)),...,\varrho^1(\gamma_N(t))) \quad 0\leq t \leq N \\
   &(\varrho^2(1-\gamma_1(t)),...,1-\varrho^2(\gamma_N(t))) \quad 0\leq t \leq N.
\end{aligned}
\end{equation}
Then, Theorem \ref{theo4} is introduced to define the expected repair cost of the repair operations of the network in a period. 
\begin{theorem}\label{theo4}
 The expected amount of cost in the time horizon $[0, T]$ is
\begin{equation}\label{eq_cost}
\begin{aligned}
   C(\bm{w}) =&\int_0^T\sum_{i=1}^N\phi_i(\delta_i(t))S_i(t,\bm{w}(t),G(t))dt \\
   + &\int_0^T\sum_{i=1}^N\varrho^1_i(\gamma_i(t))[1-M_i(t,\bm{w}(t),G(t))\\-&S_i(t,\bm{w}(t),G(t))-H_i(t,\bm{w}(t),G(t))]dt \\
   + &\int_0^T\sum_{i=1}^N\varrho^2_i(1-\gamma_i(t))[1-M_i(t,\bm{w}(t),G(t))\\-&S_i(t,\bm{w}(t),G(t))-H_i(t,\bm{w}(t),G(t))]dt.
   \end{aligned}
\end{equation}
\end{theorem}
\begin{IEEEproof}
The proof of the cost is comparable to the proof in Theorem \ref{theo2}, and we omit it here.
\end{IEEEproof}
\begin{remark}
Through the quarantine cost function $\bm{\phi}$ with recovery cost function $\bm{\varrho^1}$ and $\bm{\varrho^2}$, we define the expected cost of the repair operations in the infinitesimal time of one node. Based on the infinitesimal cost of each node, it is easy to represent the expected cost of the repair operations in a time period by the integral formulation.
\end{remark}
\par
   According to Theorems \ref{theo2}, \ref{theo3}, and \ref{theo4}, the total expected impact is
\begin{equation}\label{eq_totalimpact}
\begin{aligned}
I(\bm{w}) = &L(\bm{w})+E(\bm{w})+C(\bm{w}) \\
=&\int_0^TI(\bm{S}(t,\bm{w}(t)),\bm{w}(t),G(t))dt,\\
   \end{aligned}
\end{equation}
where 
\begin{equation}\label{eq_total impact}
\begin{aligned}
&I(\bm{E}(t,\bm{w}(t)),\bm{w}(t))=\\ &\int_0^T\sum_{i=1}^N[(\phi_i(\delta_i(t))+a^2_i)S_i(t)+a^1_iM_i(t)]dt\\
+&\int_0^T\sum_{i=1}^N [b_i+\varrho^1_i(\gamma_i(t))+\varrho^2_i(1-\gamma_i(t)]*\\
&[1-M_i(t)-S_i(t)-H_i(t)]dt.
   \end{aligned}
\end{equation}

\section{Dynamic APT Repair Game Model and PPAC Problem Formulation}\label{PPAC formulation}
 
In this section, we first model the interactions between defenders and attackers as a dynamic APT repair game. Then, we formulate the dynamic APT repair game as a PPAC problem targeted to minimize the defender's potential loss.
\subsection{Modelling the Dynamic APT Repair Game}
In this subsection, we first analyze the subgame, i.e., the dynamic APT repair game in one time slot of the ATP attack process as below.
\subsubsection{Subgame Modelling}
In an APT attack process slot, defenders deploy defensive measures far-sighted to minimize the organization's loss. We model the behaviors of attackers and defenders in each time slot as a repair game as follows:
\begin{game}\label{game1}
(Repair Game in an APT Attack Slot).
\begin{itemize}
      \item $Players$: each host $v_i \in A_j$ of the organization.
      \item $Strategies$: depending on each node's average rate $\delta_i \in \{\underline{\delta_{i}}, \overline{{\delta}_{i}}\}$, $\lambda_i \in \{\underline{\lambda_{i}}, \overline{\lambda}_{i}\}$, and $\gamma_i \{\underline{\gamma_{i}}, \overline{\gamma_{i}}\}$ defined in Eq. (\ref{eq_differential dyna sys}), defenders deploy the different defense measures at each node $v_i$ defined in Eq. (\ref{eq_w}) for organization loss reducing.
      \item $Objectives$: defenders aim to minimize the total impact of APT defined in Eq (\ref{eq_totalimpact}).
\end{itemize}
\begin{remark}
\label{r5}
In this paper, we assume the attacker adopts the greedy strategy, i.e., maximizing the loss to the target organization in the current game. The reason behind the attacker's inclination towards a greedy strategy stems from the inherent information asymmetry, where the attacker lacks comprehensive access to the organization's global information, i.e., network topology, access control settings, asset distribution, etc. Limited to local information, such as the organization's network segment or specific service versions, the attacker strives to maximize immediate gains in each operation. According to \cite{ye2020differentially}, under the circumstances of information asymmetry, it is considered a judicious course of action for attackers to adopt a greedy strategy.
\end{remark}
\end{game}
\par
 Each element of APT process sequence Eq. (\ref{eq_T}) denotes the APT attack time slot of the subgame, such as $A_i=(t_j,t_{j+1})$ represents $i-th$ subgame and $[t_j,t_{j+1}]$ represents the time horizon of this subgame. Next, we model the defenders’ resource allocation behaviors as a sequential game in the APT attack process.
\subsubsection{Sequential Game Modelling}
Before modeling the sequential game, we introduce the interactions between the subgame. Defenders are often passive \cite{husak2021predictive} in an APT defense process because defenders have no grounds to deny users access to the network services. The defenders' sensible choice is to respond quickly after observing the attack. Therefore in a subgame, the attacker first determines its attack strategy in Stage I, and then the defender determines its defense strategy in Stage II as a response to the observed behaviors of the attacker. These two stages repeat in the subgame, and the strategies of this subgame are known for the next subgame. The sequential game to model the interactions among attacks and defenders in the finite time horizon is as follows.
\begin{game}\label{game2}
(Sequential Game in APT Attack Process).
\begin{itemize}
        \item $Players$: each host $v_i \in A_j$ of the organization.
      \item $Strategies$: average rate vector $\bm{w(t)}$ defined in Eq. (\ref{eq_differential dyna sys}) in each time slot t.
      \item $Histories$: the action of previous subgame $\bm{w(t-1)}$.
      \item $Objectives$: defenders aim to minimize the total impact of APT in the whole APT attack process.
\end{itemize}
\end{game}
\begin{remark}
In the above sequential game, defenders decide the defense strategy timely to minimize the loss of the entire APT attack process.
\end{remark}
\par
 \subsubsection{Game Analysis}
As demonstrated in Eq. (\ref{eq_trans}), the attacker can manipulate the selection of target hosts by controlling the parameters $\alpha$ and $\beta$, and can adjust the intensity of attacks on the target hosts by tuning the parameter $\lambda$. We define $(PC[0,T])^{k}$ as the set of all piecewise continuous k-dimensional functions defined on the interval $[0, T]$.\par
Define a $(N)$-dimensional function $\mathbf{A_t}$ as follows: 
\begin{equation}
\label{AttackStrategy}
\begin{split}
\mathbf{A_t}=(\lambda_1(t),...,\lambda_N(t))  \quad 0\leq t \leq N
\end{split}
\end{equation}
Then the strategy of the attacker can be represented as $\mathbb{S}_A=\mathbf{A_t} \in (PC[0,T])^{N}$. \par
In contrast, defenders have the ability to optimize host maintenance strategies by controlling the parameters $\delta$ and $\gamma$. Define a $(N+N)$-dimensional function $\mathbf{D}_f$ as follows:
\begin{equation}
\label{DefenderStrategy}
\begin{split}
\mathbf{D}_f=(\delta_1(t),...,\delta_N(t),...,\gamma_1(t),...,\gamma_N(t))  \quad 0\leq t \leq N .
\end{split}
\end{equation}
The strategy of defender can be represented as $\mathbb{S}_D=\mathbf{D}_f \in (PC[0,T])^{N+N}$. 
After defining the strategy sets for the attacker and defender, we assume that the strategy spaces for both can be partitioned into discrete sets \cite{myerson1991game}. Let the strategy set for attackers have $m$ options, denoted as $\mathbb{S}_A=\{ \mathbf{A}^t_1,...,\mathbf{A}^t_m\}$. Similarly, the strategy set for defenders has $n$ options, denoted as $\mathbb{S}_D=\{ \mathbf{D}^f_1,...,\mathbf{D}^f_n\}$. The decisions of attackers and defenders form the strategy pairs $(\mathbf{A}^t_i,\mathbf{D}^f_j)$. Naturally, there are a total of $m \times n$ strategy pairs in this situation. Assuming that the winning strategy for the defender is denoted as $\mathbf{A}^t_{ij}$, we can obtain the matrix $\mathbf{A}$ in Eq. (\ref{matrix1}).
\begin{equation}
\begin{bmatrix}
\label{matrix1}
\mathbf{A}^t_{11} & \mathbf{A}^t_{12} & \dots & \mathbf{A}^t_{1n} \\
\mathbf{A}^t_{21} & \mathbf{A}^t_{22} & \dots & \mathbf{A}^t_{2n} \\
\vdots & \vdots & \ddots & \vdots \\
\mathbf{A}^t_{m1} & \mathbf{A}^t_{m2} & \dots & \mathbf{A}^t_{mn} \\
\end{bmatrix}
\end{equation}
Based on the analysis of attack impact in Sect.~III.C, the attacker's benefit is determined as follows: 
\begin{equation}
\label{attackbenefit}
\begin{split}
\mathbf{A_b}= \mathbf{W^{A}}\cdot I(\bm{E}(t,\bm{w}(t)),\bm{w}(t)).
\end{split}
\end{equation}
The defender's benefit is determined as
\begin{equation}
\label{defenderbenefit}
\begin{split}
\mathbf{D_b}= \mathbf{W^{D}}\cdot I(\bm{E}(t,\bm{w}(t)),\bm{w}(t)).
\end{split}
\end{equation}
The $\mathbf{W^{A}}$ and $\mathbf{W^{D}}$ represent the coefficients of impact of attacks and defenders, where $\mathbf{W^{A}} \textgreater{} 0$ and $\mathbf{W^{D}} \textless{} 0$. The impact of attacks is a positive gain for attackers, but a loss for defenders. It should be noted that the matrix $\mathbf{A}$ also represents the loss matrix for the attacker, thus the payoff matrix for the attacker is $-\mathbf{A}$. In the process of the game, it is prudent to consider the opponent's motivation to maximize one's own loss (given the intense competition). Therefore, one should strive for the best possible outcome in the worst-case scenario. In mathematical terms, this corresponds to first taking the minimum of each row element (assuming the opponent minimizes one's loss), and then taking the maximum of all the minimum values across rows (resulting in the minimum loss), such as  
\begin{equation}
\label{sum-zero}
\begin{split}
I(\bm{E}(t,(\mathbb{S}_{A}^*,\mathbb{S}_{D}^*),(\mathbb{S}_{A}^*,\mathbb{S}_{D}^*))) = 
\max_{i} \min_{j} \mathbf{A}^t_{mn}.
\end{split}
\end{equation}
In such a zero-sum game, a strategy pair $(\mathbb{S}_{A}^*, \mathbb{S}_{D}^*) \in \mathbf{A_t} \times \mathbf{D}_f$ is referred to as a Nash equilibrium if 
\begin{equation}
\label{nash equilibrium}
\begin{split}
I(\bm{E}(t,(\mathbb{S}_{A}^*,\mathbb{S}_{D}^*),(\mathbb{S}_{A}^*,\mathbb{S}_{D}^*)) \geq I(\bm{E}(t,(\mathbb{S}_{A},\mathbb{S}_{D}),(\mathbb{S}_{A},\mathbb{S}_{D})).
\end{split}
\end{equation}
\par
Once the subgame equilibrium has been determined, we can then proceed to analyze the equilibrium of the entire sequential game. Considering the strategic interactions and decision-making of the players across all stages of the game, we aim to identify the strategies and outcomes that constitute an equilibrium for the overall sequential game. This entails examining the consistency and optimality of player strategies, taking into account the sequential nature of the game and the information available to each player at different stages.
\begin{theorem}\label{themo_6}
Dynamic games in which players have a finite set of actions and engage in a finite number of interactions exhibit equivalence, while every finite extensive-form game characterized by perfect information possesses a pure-strategy Nash equilibrium \cite{deng2016adaptive}.
\end{theorem}
\begin{IEEEproof}
The proof for Theorem 6 is provided in \cite{kuhn11953extensive}.
\end{IEEEproof}
Theorem \ref{themo_6} shows the existence of a Nash equilibrium in the formulated sequential game. Moreover, it guarantees the convergence of the game to this equilibrium within a finite number of iterations. Hence, the equilibrium of the game is $\mathbb{S}=((\mathbb{S}_{A}^*,\mathbb{S}_{D}^*)^1,...,(\mathbb{S}_{A}^*,\mathbb{S}_{D}^*)^N).$
\begin{figure*}[!t]
\centering
\begin{center}
\includegraphics[width=\linewidth]{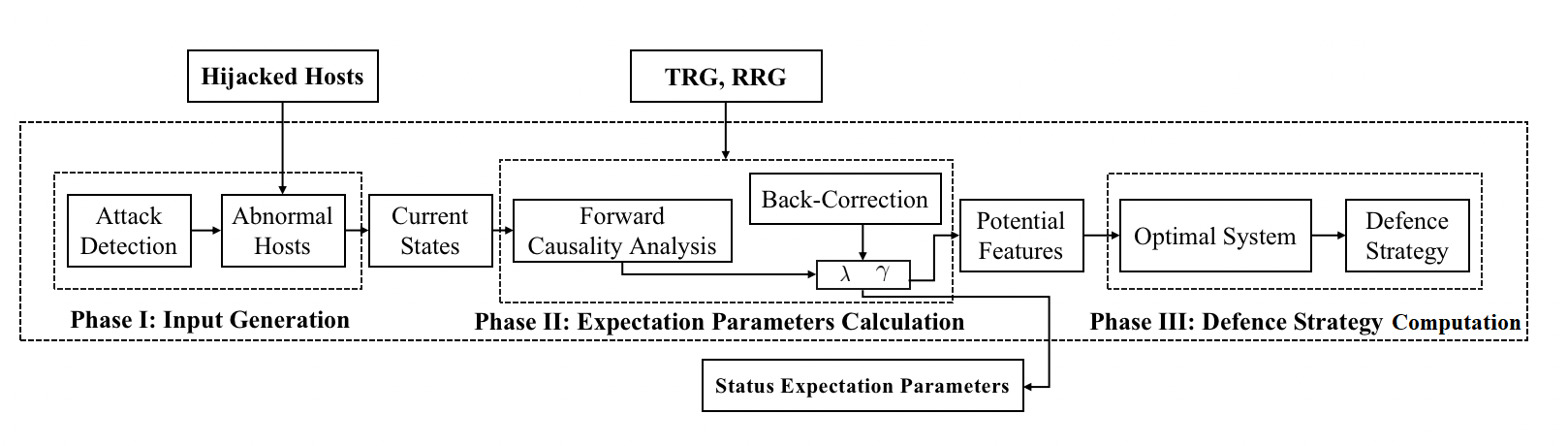}
\caption{Architecture of the proposed scheme}
\label{overview}
\end{center}
\end{figure*}
\subsection{PPAC Problem Formulation}
In this subsection, we formulate the PPAC Problem based on the game model. Firstly, we assume the solution function of PPAC is piecewise continuous in this paper. Hence, we define $PC[0,T]$ as the function space of the attack process, and the PPAC's solution set is $\bm{w}\in (PC[0,T])^{3N}$. Secondly, we assume all quarantine rate functions, all threat rate functions, and all recovery rate functions in each feasible strategy are bounded by the dynamic defense resource. Thereby, for $1\leq i \leq N$, $\underline{\delta_i}, \; \overline{\delta_i}$ represent upper and lower bounds of $\delta_i$, $\underline{\lambda_i}, \; \overline{\lambda_i}$ represent upper and lower bounds of $\lambda_i$, and $\underline{\gamma_i}, \; \overline{\gamma_i}$ represent upper and lower bounds of $\gamma_i$. For brevity, we denote $\underline{\bm{\delta}}=(\underline{\delta_1},...,\underline{\delta_N}),$ $\overline{\bm{\delta}}=(\overline{\delta_1},...,\overline{\delta_N}),$ $\underline{\bm{\lambda}}=(\underline{\lambda_1},...,\underline{\lambda_N}),$ $\overline{\bm{\lambda}}=(\overline{\lambda_1},...,\overline{\lambda_N}),$ $\underline{\bm{\gamma}}=(\underline{\gamma_1},...,\underline{\gamma_N}),$ $\overline{\bm{\gamma}}=(\overline{\gamma_1},...,\overline{\gamma_N})$. Note that the value of $\underline{\bm{\delta}},$ $\overline{\bm{\delta}},$ $\underline{\bm{\lambda}},$ $\overline{\bm{\lambda}},$ $\underline{\bm{\gamma}}$, and $\overline{\bm{\gamma}}$ can be estimated by the previous data and the experience of defenders for different scenarios. Finally, the feasible solution set of the PPAC problem can be summarized as 
\begin{equation}\label{eq_set of PPAC}
\begin{aligned}
\mathcal{W}=\{&\bm{w}\in (PC[0,T])^{3N}|\bm{w}(t)\in \\
&\prod^N_{i=1}[\underline{\delta_i},\overline{\delta_i}]\times\prod^N_{i=1}[\underline{\lambda_i},\overline{\lambda_i}]\times\prod^N_{i=1}[\underline{\gamma_i},\overline{\gamma_i}] \}.
   \end{aligned}
\end{equation}
\par
 After defining the above basic variables, we proceed with formulating the PPAC problem. To better align the problem with real-world attack-defense scenarios, our work is based on three practical assumptions: (a) We assume that the entire attack-defense process can be divided into multiple sub-stages based on time \cite{wrightson2014advanced}, and the defender has access to the current stage's global network information and all the information from the previous round, while the attacker can only access the information from the previous round. (b) In each round of the process, the attacker takes action first, and due to the unequal information between the attacker and defender, the attacker adopts a greedy strategy (further details can be found in Remark \ref{r5}), pursuing maximum immediate gains without considering the impact on the next stage. On the other hand, the defender leverages their advantage of having more information and considers the impact of their belief about the attacker's decision in the next round \cite{dawkins2004systematic}. (c) At time period $t+1$, the defender's relative belief is solely determined by the historical information and the defender's action at time period $t$ \cite{fudenberg1991game}. 
\par
Define vector $\mu = (\mu^1,...,\mu^m)^T$ as the belief vector of defenders, and vector $\rho = (\rho^1,...,\rho^m)^T$ as the belief vector of attackers. Then the expectation of defenders win is $\mathbb{E}(\mathbb{S}_A,\mathbb{S}_D)=\mu^T \mathbf{A} \rho$. Then, the objective from defender's perspective can be expressed as 
\begin{equation}
\begin{aligned}
 \max_{\mu} \min_{\rho}\mu^T \mathbf{A} (\sum^n_{j=1}\rho_je_j),
   \end{aligned}
\end{equation}
where $e_j$ represents a vector with only the $j-th$ component being $1$ and all other components being $0$. Therefore, from the defender's perspective and the aforementioned assumptions, we can formulate the optimization problem for the defenders as 
\begin{equation}
\begin{aligned}
 \min I(\bm{w})= \max_{\mu}(\sum^n_{j=1}\rho_je_j).
\end{aligned}
\end{equation}
The optimization problem for the attackers is 
\begin{equation}
\begin{aligned}
 \max -I(\bm{w})= \min_{\rho}(\sum^n_{i=1}\mu_ie_i).
\end{aligned}
\end{equation}
\par
Therefore, the decision problems of the attacker and defender are dual linear problems and have the same optimal objective function value. Hence, finding the optimal solution for either of them would suffice. To capture the influence of attacker's decisions on defender's decisions, we enhance the optimization problem for the defender by incorporating the factors of attacker's decisions into defender's objective function. Therefore, for the PPAC problem, a decision can be represented as $\bm{w}(t)$. Hence, the PPAC problem can be formulated as  
\begin{equation}\label{eq_problem}
\begin{aligned}
&\min_{\substack{\bm{w}\in\mathcal{W}}}\quad I(\bm{w}) =\int_0^TI(\bm{E}(t,\bm{w}(t)),\bm{w}(t),G(t))dt,\\
&s.t. 
 \begin{cases}
     &\frac{d\bm{E}(t)}{dt}=\bm{F}(\bm{E}(t),\bm{w}(t))  \\
     &\lambda_i(t)=T(v_i,t) \\
     &\gamma_i(t)=R(v_i,t)\\
     &\bm{E}(0)=\bm{E}_0 \\
     &\bm{U_n}\in(0,\bm{U})
 \end{cases}
   \end{aligned}
\end{equation}
 where $\frac{d\bm{E}(t)}{dt}=\bm{F}(\bm{E}(t),\bm{w}(t))$ is a brief expression of Eq. (\ref{eq_differential dyna sys}). $T(v_i,t)$ and $R(v_i,t)$ represent the node $v_i$ in the threat algorithm TRG and recovery algorithm RRG, respectively. $\bm{U}$ is the utility of the network work without APT attack, and $\bm{U_n}$ is the minimum utility of the task need.\par
\begin{remark}
The admissible control set of the optimal problem in Eq. (\ref{eq_problem}) represents the feasible defense strategies of the dynamic APT repair game. The objective function means the APT impact of the organization under a workable sequence, and each optimal solution denotes potential PPAC defense strategies that minimize the expected APT impact. 
\end{remark}
\par
The PPAC problem can be summarized as 
\begin{equation}\label{eq_PPAC}
\begin{aligned}
\mathbb{M}=(G,a^1_i,a^2_i,b_i,\phi,\varrho^1,\varrho^2,\alpha,\beta,\delta,\lambda,\gamma,\bm{\hat{T}},R,T,\bm{U_n},\bm{U},\bm{E}_0),
   \end{aligned}
\end{equation}
where $\lambda$ and $\gamma$ are parameters determined by algorithm TRG and algorithm RRG, respectively.
\section{SOLUTION TO THE PPAC PROBLEM}

\subsection{Solution Overview}\label{view}
Fig. \ref{overview} shows the workflow of our proposed game solution to design optimal defense strategies under an APT alert (i.e., the current anomalous hosts). As previously analyzed, the dynamic variation of the network and the impact of attack and defense interactions make the solution strategy easy to fall into a locally optimal solution. To escape the local optimal solution trap, we develop two backtracking-forward rating algorithms for risk assessment in phase II. For timely defense, we build an optimal system in phase III. The detail of the three phases in our solution is shown below.
\begin{itemize}
    \item $Phase\enspace I$: In this phase, the attacker obtains the anomalous hosts in the current network based on the attack detection methods \cite{fang2022back,alsaheel2021atlas,tuor2017deep,milajerdi2019holmes,zengy2022shadewatcher} and uses these anomalous hosts as input for the next phase.  
    \item $Phase\enspace II$: Depending on phase I, the TRG and RRG compute the status expectation parameters $\lambda$ and $\gamma$, respectively. The details are shown in Section \ref{algorithm}.
    \item $Phase\enspace III$: We construct a dynamic system based on the solution objective of the PPAC problem. To simplify the solving, we extend the solving set to segmented functions through Pontryagin maximum principle. Based on the above and the parameters of phase II, phase III can give the optimal strategies to solve the PPAC problem. The detail of phase III is shown in Section \ref{opt sys}.
\end{itemize}

\subsection{Backtracking-Forward Rating Algorithm} \label{algorithm}
The time-varying network topology leads to the optimal behavior in the current network state becoming a locally optimal solution in the global aspect. Although this problem can be solved by increasing the dynamic inputs to the optimal system, the system with various dynamic inputs will make the interactions between inputs complex. To solve this problem, we develop two backtracking-forward rating algorithms, TRG and RRG. The three key factors and the corresponding algorithm operations for the classification of a node are described as follows.

 \begin{itemize}
        \item \textbf{Structural features}: The importance of the topological relationship between nodes in the network. Some nodes in the network are critical to its integrity due to their relative position, degree, and the betweenness \cite{freeman1977set}. To represent the topological importance of a node on the network, we use the k-shell \cite{maji2020influential} algorithm to characterize the range of node's influence on the network and use the betweenness to characterize the influence of a node on the connectivity of the network.
        \item \textbf{Lateral features, which is also called infection features}: The node's importance to the state matrix's spectral radius after a lateral attack on the entire network. For example, if a service host in an organization with an open public port is attacked, the attacker can use the machine as a broiler to attack devices that require a specific IP to open the port. Our algorithms use epidemic thresholds \cite{li2021dynamics} for node lateral attack descriptions. We use the spectral radius of the entire state matrix as described in the \cite{zino2021analysis} to discuss epidemic thresholds.
        \item \textbf{Service features}: The importance of nodes to the service performance of the entire network. The impact of the node itself on the overall service function is determined by two main aspects: its original service capability and its role in the network for information transfer. Service utility information is readily available to defenders. In this paper, we take in the experiment according to the topology of randomly generated service weights to represent his service capabilities.
    \end{itemize}
\par

  Note that, the spectral radius in this paper is $ \rho(\boldsymbol{G})=\max _{1 \leq i \leq n}\left|\lambda_{i}\right|$, where $\rho(\Lambda A-M)$ indicates the epidemic threshold. In $\rho(\Lambda A-M)$, $\Lambda$ represents the infection rate diagonal matrix, $M$ represents the recovery rate diagonal matrix, and $A$ is the adjacency matrix of the network. 
 The utility of the overall network is $Z=N\times \overline{I}$, where the $\overline{I}$ represents the average utility of each node, and $N$ is the number of the nodes. In what follows, we illustrate the TRG algorithm and RRG algorithm in detail, respectively.
\begin{algorithm}[h]
\small
\renewcommand{\algorithmicrequire}{\textbf{Input:}}
\renewcommand{\algorithmicensure}{\textbf{Output:}}
\caption{TRG}\label{alg:alg1} 
\begin{algorithmic}[1]
\REQUIRE $t,\bm{E}(t),\hat{\bm{T}},\bm{U},\bm{U_n},\overline{I},k,t_w$
\ENSURE  the threaten state list $M$
\STATE $N=|\hat{\bm{T}}|$, which is the elements number of $\hat{\bm{T}}$;
\STATE $G=\hat{\bm{T}}(t)=(V,E)$, $V=(v_1,...,v_n)$;
\FOR{$1 \leq i \leq n$}
  \STATE $P_{v_i}$: Compromise rate of neighbor nodes within one hop of aggregation node $v_i$ by Eq. (\ref{eq_trans}) and $t_w$;
  \ENDFOR
 \FOR{$P_{v_i} \in \{P_{v_1},...,P_{v_n}\}$}
    \WHILE{$k \leq N$}
        \STATE $G=\hat{\bm{T}}(k)$;
        \STATE $P_{v_i}=P_{v_i}+\rho_{\phi}\rho_{\epsilon}+Z_{\phi}Z_{\epsilon}$;
        \STATE Compute $Z^{(k)}$, which is the overall utility of the entire network after removing node $v_i$ from the network;
        \STATE Compute $\rho^{(k)}$, which is the spectral radius at moment $t$ after removing node $v_i$ from the network;
        \IF{[$\rho_{\phi}(\rho^{(k)}-\rho^{(k-1)})-Z_{\phi}(Z^{(k)}-Z^{(k-1)})]>0$}
        \STATE $k++$;
        \ELSE 
         \STATE $\rho_{\epsilon}=\rho_{\epsilon}+\rho^{(k)}-\rho^{(k-1)}$;
         \STATE $Z_{\epsilon}=Z_{\epsilon}+Z^{(k)}-Z^{(k-1)}$;
         \STATE $k--$;
         \ENDIF
    \ENDWHILE
 \ENDFOR
 \STATE sorting $V$ in descending order;
 \STATE $j=\frac{(\bm{U}-\bm{U_n})}{\overline{I}}$;
 \STATE $P=P_{v_1}+,...,+P_{v_j}$;
 \FOR{$1\leq i \leq n$}
  \STATE $M_i=\frac{p_{v_i}}{P}$;
 \ENDFOR\
 \STATE return(T).
\end{algorithmic}
\end{algorithm}
\par
The computational complexity of Algorithm 1 is $O(n\cdot N^2)$, where $n$ and $N$ represent the number of nodes and the number of periods, respectively. In line 6 of Algorithm 1, $n$ nodes are performing utility calculations. This calculation involves a nested loop, where the number of iterations is $n\cdot N$. In lines 12-18 of algorithm 1, we can observe that in the worst case, the entire nested loop will backtrack to the initial position after each iteration. Thus, the number of worst iterations in this loop is $\frac{N(1+N)}{2}$. The overall number of iterations is $n\cdot \frac{N(1+N)}{2}$. 
\subsubsection{The TRG Algorithm}
The TRG algorithm is used to classify the hazard level. The dynamic contagion problem is summarized in \cite{zino2021analysis}, which is proved to be NP-hard under the edge-level model. TRG adopts the node-level method to solve the dynamic contagion problem. \par
The incoming time $t$ and sequence $\hat{\bm{T}}$ are used to determine the current position $k$ in the time. Then, we adopt the k-shell algorithm to get each node's threat weight $ks$, which can help us better quantify each node's importance in the network. The k-shell algorithm obtains the $ks$ coefficient of each node, and the $ks$ coefficient is defined as the threat weight of each node. We denote $\overline{I}$ representing each host's average service performance.\par
 
 The algorithm starts by ranking the utility of the systems with initializing the input sequence lines 1-3. Next, we depend on the Eq. (\ref{eq_trans}) and the threat weight to aggregate the compromise rate of neighbors within one hop of each node which we define as $p_{v_i}$, lines 4. Then, we take each node with its $p_{v_i}$ to determine the threat factor. We measure nodes' contagion by the epidemic threshold of the network after removing the node and its utility by the change in the service utility of the network after removing the node, lines 4-7. Then we rely on the topological relation transformations throughout the task as well as the variation of the spectral radius to give two coefficients $\rho_{\phi}$ and $Z_{\phi}$ to measure the impact. Furthermore, the algorithm describes the impact of the current judgment on the next epoch by $\rho_{\epsilon}$ and $Z_{\epsilon}$, lines 8-13. The primary objective is to ascertain the adherence of the preceding period's decision to the prescribed criteria. Suppose we find an increase in contagiousness or a decrease in utility. In that case, we backtrack to the previous epoch with this information until a node has completed all epochs to get its final value $p_{v_i}$, lines 14-20. 
\par
\begin{figure*}[t]
 \begin{small}   
    \begin{equation}
    \label{equ:Hamiltonian}
    \begin{aligned}
&\bm{H}(\bm{E},\bm{w},\bm{\kappa},\bm{\rho},\bm{\xi})= \sum_{i=1}^{N} \kappa_i \biggl[ \gamma_i(t)[1-M_i(t)-S_i(t)-H_i(t)] \biggr.\left.- H_i(t) \biggl(\beta_i(t)+\alpha_i\sum_{j=1}^na_{ij}(t)M_j(t)+\alpha_i\sum_{j=1}^na_{ij}(t)S_j(t)\right) \biggr]\\ & +I(\bm{E}(t),\bm{w}(t))+\sum_{i=1}^{N} \xi_i(\lambda_i(t)M_i(t)-\delta_i(t)S_i(t))+\sum_{i=1}^{N} \rho_i\biggl[H_i(t)\biggl(\beta_i(t)+\alpha_i\sum_{j=1}^na_{ij}(t)M_j(t)
+\alpha_i\sum_{j=1}^na_{ij}(t)S_j(t)\biggl) \biggr.
 \biggr.-\lambda_i(t)M_i(t)\biggr]
   \end{aligned}  
    \end{equation}
    \end{small}
    \hrulefill
\end{figure*}
These values are propagated and assimilated into the network model, facilitating the evaluation and validation of the updated network topology during this period. The primary objective is to ascertain the adherence of the preceding period's decision to the prescribed criteria. If the decision is deemed compliant, the progression continues to the next period. However, if discrepancies are detected, and the decision fails to meet the stipulated requirements, the disparity value $\rho_{\epsilon}$ and $Z_{\epsilon}$ are utilized as an adjacent solution to rectify the decision made in the previous stage.\par
 To ensure the credibility of the rating, we make a worst-case assumption about the evolution of the node and find the potentially infected node with the largest structural and efficacy characteristics as the default infected node. Finally, we sort the nodes by these final values and get the threat rate by the ratio of the theoretical values. The threat rate is generated by comparing the theoretical worst total impact value at that moment with the theoretical value for each specific node, lines 20-24. The algorithm finally outputs a list of threat registration assessments $M$ for all nodes at that moment, line 24. The details of RRG algorithm are shown in Algorithm \ref{alg:alg1}.  \par
\begin{figure*}[!b]
\begin{small}
\begin{equation}\label{eq_co-state}
\begin{aligned}
\frac{d\kappa(t)}{dt}=&bi+(\varrho^1_i-\varrho^2_i+\kappa_i(t))\gamma_i+(\kappa_i(t)-\rho_i(t))  \biggl(\beta_i+\alpha_i\sum_{j=1}^na_{ij}(t)M_j(t)+\alpha_i\sum_{j=1}^na_{ij}(t)S_j(t) \biggr)\\
    \frac{d\rho(t)}{dt}=&-a^1_i+bi+\varrho^2_i+(\varrho^1_i-\varrho^2_i+\kappa_i(t))\gamma_i(t)+(\rho_i(t)-\xi_i(t))\lambda_i+[\rho_i(t)-\kappa_i(t)]\sum^N_{j=1}\alpha_ia_{ij}H_j(t)\\
    \frac{d\xi(t)}{dt}=&-a^2_i-\phi_i(\delta_i(t))+b_i+\varrho^2_i+(\varrho^1_i-\varrho^2_i+\kappa_i)\gamma_i+\xi_i(t)\delta_i(t)+[\rho_i(t)-\kappa_i(t)]\sum^N_{j=1}\alpha_ia_{ij}H_j(t).
   \end{aligned}
\end{equation}
\end{small}
\hrulefill
\vspace*{4pt}
\end{figure*}
\begin{algorithm}[h]
\small
\renewcommand{\algorithmicrequire}{\textbf{Input:}}
\renewcommand{\algorithmicensure}{\textbf{Output:}}
\caption{RRG}\label{alg:alg2} 
\begin{algorithmic}[1]
\REQUIRE $t,\bm{E}(t),\hat{\bm{T}},\bm{U},\bm{U_n},\overline{I},k,t_w$ 
\ENSURE  the recovery state list $M$
\STATE $N=|\hat{\bm{T}}|$, which is the elements number of $\hat{\bm{T}}$.
\STATE $G=\hat{\bm{T}}(t)=(V,E)$
\FOR{$1 \leq i \leq n$}
  \STATE $P_{v_i}$: Compromise rate of neighbor nodes within one hop of aggregation node $v_i$ by Eq. (\ref{eq_trans}) and $t_w$;
  \ENDFOR
 \FOR{$P_{v_i} \in \{P_{v_1},...,P_{v_n}\}$}
    \WHILE{$k \leq N$}
        \STATE $G=\hat{\bm{T}}(k)$;
        \STATE $P_{v_i}=P_{v_i}+\rho_{\phi}\rho_{\epsilon}+Z_{\phi}Z_{\epsilon}$;
        \STATE Compute $Z^{(k)}$, which is the overall utility of the entire network after adding node $v_i$ to the network;
        \STATE Compute $\rho^{(k)}$, which is the spectral radius at moment $t$ after adding node $v_i$ to the network;
        \IF{($Z_{\phi}(Z^{(k)}-Z^{(k-1)})-\rho_{\phi}(\rho^{(k)}-\rho^{(k-1)})]>0$}
        \STATE $k++$;
        \ELSE 
         \STATE $\rho_{\epsilon}=\rho_{\epsilon}+\rho^{(k)}-\rho^{(k-1)}$;
         \STATE $Z_{\epsilon}=Z_{\epsilon}+Z^{(k)}-Z^{(k-1)}$;
         \STATE $k--$;
         \ENDIF
    \ENDWHILE
 \ENDFOR
 \STATE sorting $V$ in descending order;
 \STATE $P=P_{v_1}+,...,+P_{v_n}$;
 \FOR{$1\leq i \leq n$}
  \STATE $M_i=\frac{p_{v_i}}{P}$;
 \ENDFOR\
 \STATE return(R).
\end{algorithmic}
\end{algorithm}
\subsubsection{The RRG Algorithm}
Due to the RRG algorithm having the same principle as TRG, we mainly introduce the difference between them. TRG and RRG are different in inputs; the inputs of RRG are the $quarantined$ nodes. RRG considers the contagion and efficiency impact of rejoining the network for the nodes separated from the network in the quarantine zone. Therefore, the RRG algorithm's back-tracing is determined by whether the effectiveness expectation is greater or smaller than the contagion expectation after the node is joined. The detail of RRG is shown in Algorithm \ref{alg:alg2}.
The computational complexity of Algorithm 2 is $O(n\cdot N^2)$, where $n$ and $N$ represent the number of nodes and the number of periods, respectively. The analysis procedure follows a similar methodology as described in Algorithm 1, and for the sake of conciseness, we omit its detailed exposition in this section.


\subsection{Optimal Control-based Defensive Strategy-making }\label{opt sys}
 For the accuracy and feasibility of the defense strategy, we denote the Eq. (\ref{eq_PPAC}) as a $3_{rd}$ order dynamic system \cite{bertsekas2019reinforcement}. In this system, the state transition parameters of the lateral movement attack model are used as the inputs. \par
 Following the optimal theory \cite{bertsekas2019reinforcement}, the Hamiltonian of PPAC problem Eq. (\ref{eq_problem}) is summarized as Eq. (\ref{equ:Hamiltonian}), where $\bm{\kappa}=(\kappa_1,...,\kappa_N)$, $\bm{\rho}=(\rho_1,...,\rho_N)$ and $\bm{\xi}=(\xi_1,...,\xi_N)$ are the associated adjoints. Then, the conditions of optimal control are proven in Theorem \ref{theo5}.

\begin{theorem}\label{theo5}
Assume that $\bm{w}$ is an optimal control of the PPAC problem, and $\bm{E}$ is the solution to the associated system (Eq. (\ref{eq_differential dyna sys})). Then the adjoint functions are $\bm{\kappa}$, $\bm{\rho}$ and $\bm{\xi}$ when $\bm{\kappa}(T)=\bm{\rho}(T)=\bm{\xi}(T)=0$. 
Furthermore, 
\begin{equation}\label{eq_co-state-further}
\left\{\begin{array}{l}
\delta_{i}(t) \in \arg \min _{\delta \in\left[\underline{\delta_{i}}, \overline{{\delta}_{i}}\right]} S_{i}(t)\left[\phi_{i}(\delta)-\xi_{i}(t) \delta\right] \\
\lambda_{i}(t) \in \arg \min _{\lambda \in\left[\underline{\lambda_{i}}, \overline{\lambda}_{i}\right]} M_{i}(t)\left[\xi_{i}(t)\lambda-\rho_{i}(t) \lambda\right] \\
\gamma_{i}(t) \in \arg \min _{\gamma \in\left[\underline{\gamma_{i}}, \overline{\gamma_{i}}\right]}\left[1-H_{i}(t)-M_{i}(t)-S_{i}(t)\right]\\
\left[\varrho^1_{i}(\gamma)+\varrho^2_{i}(1-\gamma)+\kappa_{i}(t) \gamma\right] \\

0 \leq t \leq T, i=1, \ldots, N.
\end{array}\right.
\end{equation}

\end{theorem}
\begin{IEEEproof}
According to the Pontryagin’s minimum principle, the differential equations of $\bm{\kappa}$, $\bm{\rho}$ and $\bm{\xi}$ are illustrated as  
\begin{equation}
\left\{\begin{array}{l}
\frac{d \kappa{i}(t)}{d t}=-\frac{\partial H(\mathbf{E}(t), \mathbf{w}(t), \bm{\kappa}(t), \bm{\rho}(t), \bm{\xi}(t))}{\partial H_{i}} \\
\frac{d \rho{i}(t)}{d t}=-\frac{\partial H(\mathbf{E}(t), \mathbf{w}(t), \bm{\kappa}(t), \bm{\rho}(t), \bm{\xi}(t))}{\partial M_{i}} \\
\frac{d \xi{i}(t)}{d t}=-\frac{\partial H(\mathbf{E}(t), \mathbf{w}(t), \bm{\kappa}(t), \bm{\rho}(t), \bm{\xi}(t))}{\partial S_{i}} \\
0 \leq t \leq T, i=1, \ldots, N.
\end{array}\right.
\end{equation}
Theorem \ref{theo5} is proved.
\end{IEEEproof}
\par
 Due to the terminal cost and final state are unlimited, the conditions $\bm{\kappa}(T)=\bm{\rho}(T)=\bm{\xi}(T)=0$ hold. Again, by Pontryagin's minimum principle, we have
\begin{equation}\label{eq_minw}
\mathbf{w}(t) \in \arg \min _{\widetilde{\mathbf{w}} \in \mathcal{W}} H(\mathbf{E}(t), \widetilde{\mathbf{w}}(t), \bm{\kappa}(t), \bm{\rho}(t), \bm{\xi}(t), 0 \leq t \leq T.
\end{equation}
Combining Eq. (\ref{equ:Hamiltonian}), Eq. (\ref{eq_co-state-further}), and Eq. (\ref{eq_minw}), we can deduce the Eq. (\ref{eq_H-futher}).
 \begin{figure*}[!t]
 \begin{small}
\begin{equation}\label{eq_H-futher}
\begin{aligned}
&\bm{H}(\bm{E},\widetilde{\mathbf{w}}(t),\bm{\kappa},\bm{\rho},\bm{\xi})= \int_0^T\sum_{i=1}^N[(\phi_i(\widetilde{\delta_i})(t)+a^2_i)S_i(t)+a^1_iM_i(t)]dt\int_0^T\sum_{i=1}^N [b_i+\varrho^1_i(\widetilde{\gamma_i})(t)+\varrho^2_i(1-\widetilde{\gamma_i}(t)][1-M_i(t)-S_i(t)-H_i(t)]dt\\
&+\sum_{i=1}^{N} \xi_i(\widetilde{\lambda_i}(t)M_i(t)-\widetilde{\delta_i}(t)S_i(t))+\sum_{i=1}^{N} \kappa_i \biggl[ \widetilde{\gamma_i}(t)[1-M_i(t)-S_i(t)-H_i(t)] \biggr.\left.- H_i(t) \biggl(\beta_i(t)+\alpha_i\sum_{j=1}^na_{ij}(t)M_j(t)+\alpha_i\sum_{j=1}^na_{ij}(t)S_j(t)\right)\biggr]  \\&+\sum_{i=1}^{N} \rho_i\biggl[H_i(t)\biggl(\beta_i(t)+\alpha_i\sum_{j=1}^na_{ij}(t)M_j(t)
+\alpha_i\sum_{j=1}^na_{ij}(t)S_j(t)\biggl) \biggr.
 \biggr.-\widetilde{\lambda_i}(t)M_i(t)\biggr]
   \end{aligned}
\end{equation}
\end{small}
\hrulefill
\vspace*{4pt}
\end{figure*}

According to optimal control theory, we can get the control of the PPAC problem Eq. (\ref{eq_problem}) by solving Eq. (\ref{eq_differential dyna sys}), (\ref{eq_co-state}), (\ref{eq_co-state-further}), (\ref{eq_H-futher}), and condition $\bm{\kappa}(T)=\bm{\rho}(T)=\bm{\xi}(T)=0$. We point out that the solutions of this system are potential optimal strategies of the PPAC problem. The analysis of the solutions' structure is detailed in Appendix B.\par

\section{PERFORMANCE EVALUATION} \label{experiment}

We adopt the following three experimental settings. \emph{Setting 1}: a static network (i.e., in Setting \ref{static network}); \emph{Setting 2}: a periodic time-varying network (i.e., in Setting \ref{periodic network}); \emph{Setting 3}: a time-varying network (i.e., in Setting \ref{time-varying network}). To be fair in performance comparison, we adapt well-performing techniques  ER \cite{yeDifferentiallyPrivateGame2021} (differential private game) and QAR \cite{yangEffectiveQuarantineRecovery2021} (quarantine and recovery) from static networks. We compute the results separately for each time interval and averaged them to obtain the solutions for these methods. This allows us to adapt the methods to the dynamic nature of the network and provide meaningful comparisons. \par

ER performs overall repairs without node isolation, while QAR isolates all nodes with anomalies. To ensure fairness, we compared the number of repair operations over a specific time period using the PPAC metric. In the experiments, for ER, we set the number of iterations equal to the number of nodes in the network. For QAR, we also set the number of time steps in the forward-backward method \cite{lapidus1971numerical,yangEffectiveQuarantineRecovery2021} equal to the number of nodes in the network. As such, the computational complexity of ER is $O(N\cdot n^2)$, where $n$ and $N$ represent the number of nodes and the number of periods. The computational complexity of QAR is also $O(N\cdot n^2)$, where $n$ and $N$ represent the number of nodes and the number of periods, respectively. For detailed settings of $N$ and $n$ in each experiment, please refer to the following subsections. In general, the value of $N$ is significantly smaller than $n$.\par
 We evaluate the performance of our proposed optimal strategy in terms of the following three important metrics.\par
\textbf{1) Resource occupancy}: The resource occupancy is the ratio between the resource consumption and the overall resource (i.e., the CPU, memory, human resource, etc., which we regard as equivalent to $U$) at each moment. In this paper, we assume that lower resource occupancy corresponds to better method performance. The resource consumption of each method comprises the repair resource consumption $C_r$ and the quarantine resource consumption $C_q$. Furthermore, the repair resource consumption $C_r$ correlates with the number of repair nodes $N_r$, i.e.,  $C_r \propto N_r$, and the quarantine resource consumption $C_q$ correlates with the number of repair nodes $N_q$, i.e.,  $C_q \propto N_q$. Hence, resource occupancy is $\frac{W_c^r\times\sum_{i=1}^{T}C_r+W_c^q\times\sum_{i=1}^{T}C_q}{W_c\times\frac{1}{10} \sum_{i=1}^{T}U}$, where the $W_c^r$, $W_c^q$, and $W_c$ is the consumption correction factor.\par
Different organizations often allocate varying degrees of importance to the four categories of resources: human resources, node-specific utility, network resources, and storage resources. In our experimental setup, we aim to enhance the generalizability of our results by assuming equal weights for these resource categories. For convenience, we assume the consumption correction factor of repair, quarantine, and overall resource is $W_c^r:W_c^q: W_c=1:2:4$.\footnote{Regarding resource-intensive operations, we consider that repair operations only consume human resources. For quarantine operations, both human resources and utility of the nodes are taken into account. However, no additional expenditures are required for communication and storage resources. It is worth noting that the overall resource allocation encompasses these four kinds of resources. To underscore the equitable representation of resource proportions, we introduce a coefficient that serves as a normalization factor.} \par
\textbf{2) Defense resource utilization}: For convenience, we assume the resource beyond the minimum are used for defense resources, i.e., $U-U_{n}$, and we assume that the defender expends all defensive resources to defend against the attack without strategy guidance. Hence, defense resource utilization is the ratio between the defensive resource and resource consumption,i.e., $1-\frac{W_c\times\frac{1}{10} \sum_{i=1}^{T}(U-U_{n})}{W_c^r\times\sum_{i=1}^{T}C_r+W_c^q\times\sum_{i=1}^{T}C_q}$, where $W_c\times\frac{1}{10} \sum_{i=1}^{T}(U-U_{n})$ represents the defense resource of the network. \par
\textbf{3) Service stability}: The service stability is measured by $N_s$ the number of time periods when the organization's minimum service requirements are maintained, i.e., $\frac{W_c\times\frac{1}{10} \sum_{i=1}^{T}N_s}{T}$.\par

The expression for solving the problem with the mechanism is formulated as
\begin{equation}\label{eq_PPAC-}
\begin{aligned}
\mathbb{M}=(G,\bm{a^1_i},\bm{a^2_i},\bm{b_i},\bm{\phi},\bm{\varrho^1},\bm{\varrho^2},\bm{\alpha},\bm{\beta},\bm{\underline{\delta}},\bm{\overline{\delta}},\\
\bm{\underline{\lambda}},\bm{\overline{\lambda}},\bm{\underline{\gamma}},\bm{\overline{\gamma}},\bm{\hat{T}},
R,T,\bm{U_n},\bm{U},\bm{E}_0).
   \end{aligned}
\end{equation}
\par
 Based on $\mathbb{M}$, we set the experimental parameters using the following steps. \par
 $Step$ 1: Set $\bm{a^{1*}_i}$, $\bm{a^{2*}_i}$, and $\bm{b_i}$. Define  $\bm{a^{1*}_i}$, $\bm{a^{2*}_i}$, and $\bm{b_i}^*$ as the vectors chosen randomly and uniformly from $(0, 1]^{100}$. In the following experiments, except in special cases, $\bm{a^1_i}=\bm{a^{1*}_i}$, $\bm{a^2_i}=\bm{a^{2*}_i}$, and $\bm{b_i}=\bm{b_i}^*$.\par
  $Step$ 2: Set $\bm{\alpha}$, and $\bm{\beta}$. Define $\bm{\alpha^*}=(0.1,...,0.1)$, $\bm{\beta^*}=(0.1,...,0.1)$. In our experiments, unless otherwise specified, $\bm{\alpha}=\bm{\alpha}^*$, $\bm{\beta}=\bm{\beta}^*$.\par
 $Step$ 3: Set $\bm{\underline{\delta}}$, $\bm{\overline{\delta}}$, $\bm{\underline{\lambda}}$, $\bm{\overline{\lambda}}$, $\bm{\underline{\gamma}}$, and $\bm{\overline{\gamma}}$. Define 
 \begin{equation}
	\begin{aligned}
	\bm{\underline{\delta}^c}=(0.c,...,0.c),\quad\bm{\overline{\delta}^c}=(0.c,...,0.c)\\
	\bm{\underline{\lambda}^c}=(0.c,...,0.c),\quad\bm{\overline{\lambda}^c}=(0.c,...,0.c)\\
	\bm{\underline{\gamma}^c}=(0.c,...,0.c),\quad\bm{\overline{\gamma}^c}=(0.c,...,0.c)\nonumber
	\end{aligned}
	\end{equation}
  where $c\in \{1,2,3,4,5,6\}$.\par
   $Step$ 4: Define the initial expected state $\bm{E_0}$. Let 
  \begin{equation}
	\begin{aligned}
	\bm{E}^*=(0.8,...,0.8,0.1,...,0.1,0.1,...,0.1),\nonumber
	\end{aligned}
	\end{equation}
	represent the average rate of the initial state of the node. In the following experiments, we set $\bm{E_0} = \bm{E}^*$.
  \par
  \begin{figure*}[!t]
\centering
\subfloat[]{\includegraphics[width=2.2in]{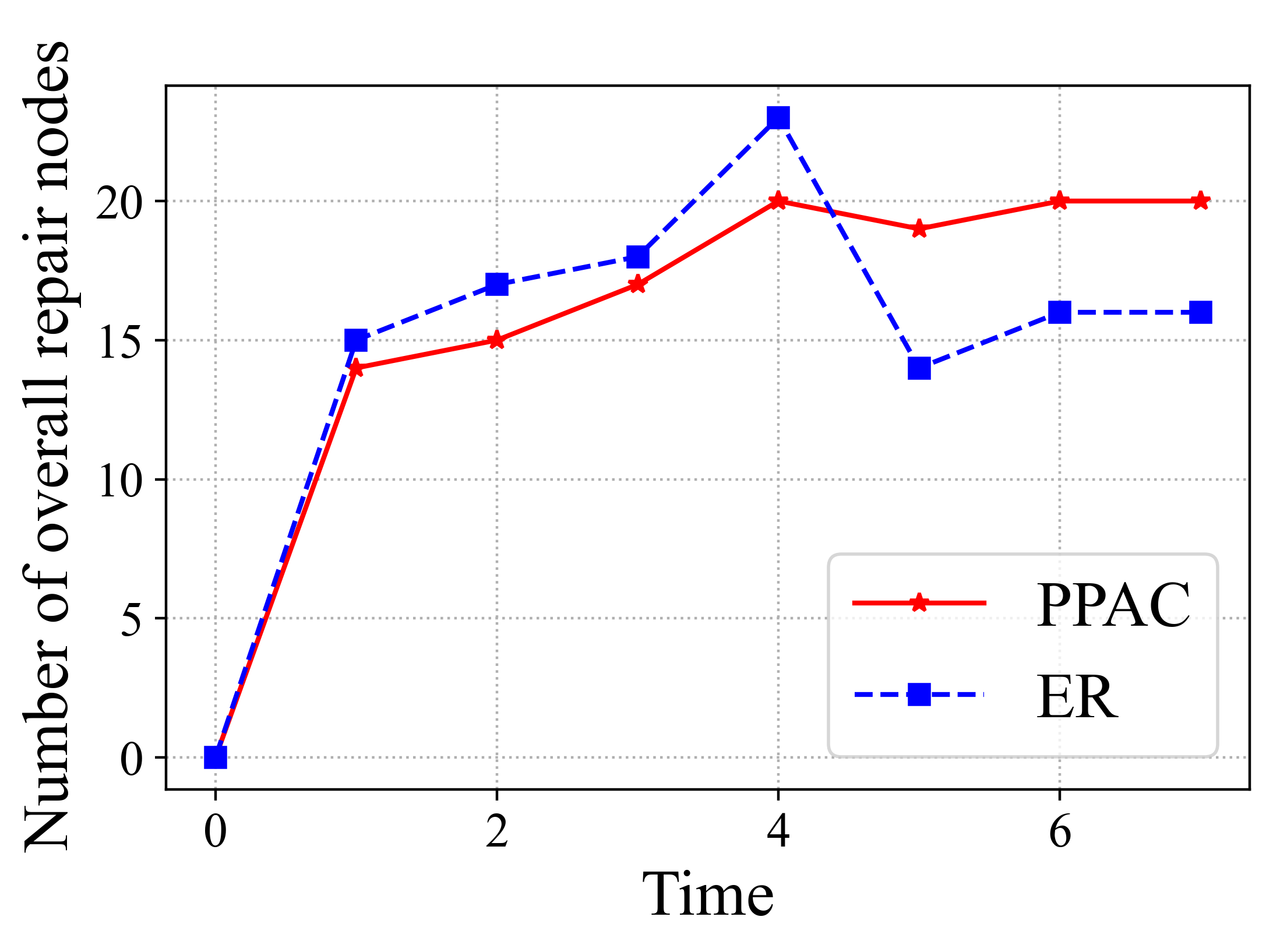}%
\label{fig_overallrepairexample3}}
\hfil
\subfloat[]{\includegraphics[width=2.2in]{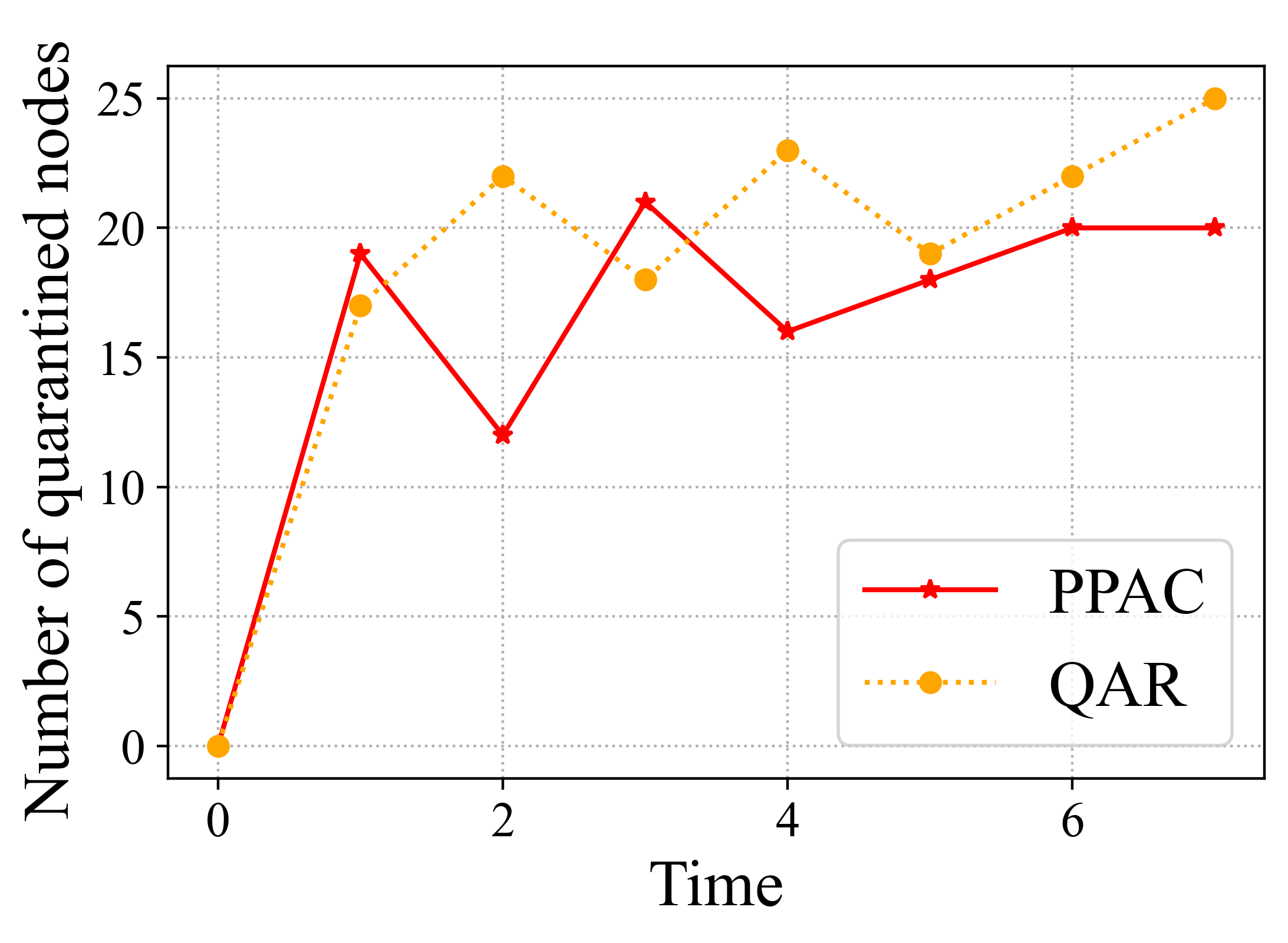}%
\label{fig_quarantinexample3}}
\hfil
\subfloat[]{\includegraphics[width=2.3in]{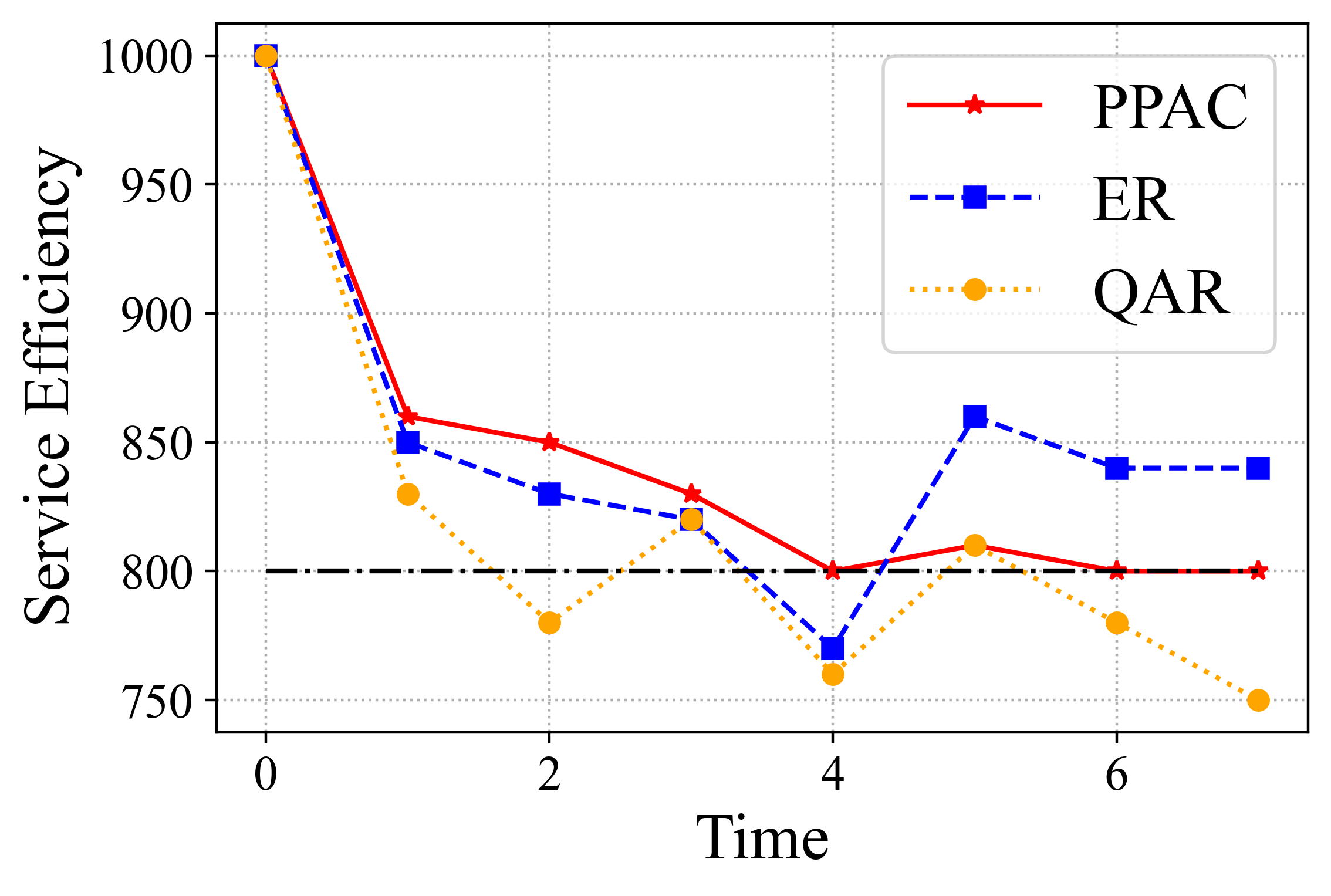}%
\label{fig_serviceultilityexample3}}
\hfil
\caption{(a) the comparison between our method and the ER by the number of nodes for overall maintenance at each moment in setting 1. (b) the comparison between our method and the QAR by the number of quarantined nodes at each moment in setting 1. (c) the comparison between our method and ER, QAR by the service quality at each time.}
\label{al}
\end{figure*} 
\begin{figure}[!h]
\centering
\subfloat[\label{fig:a}]{\includegraphics[scale=0.3]{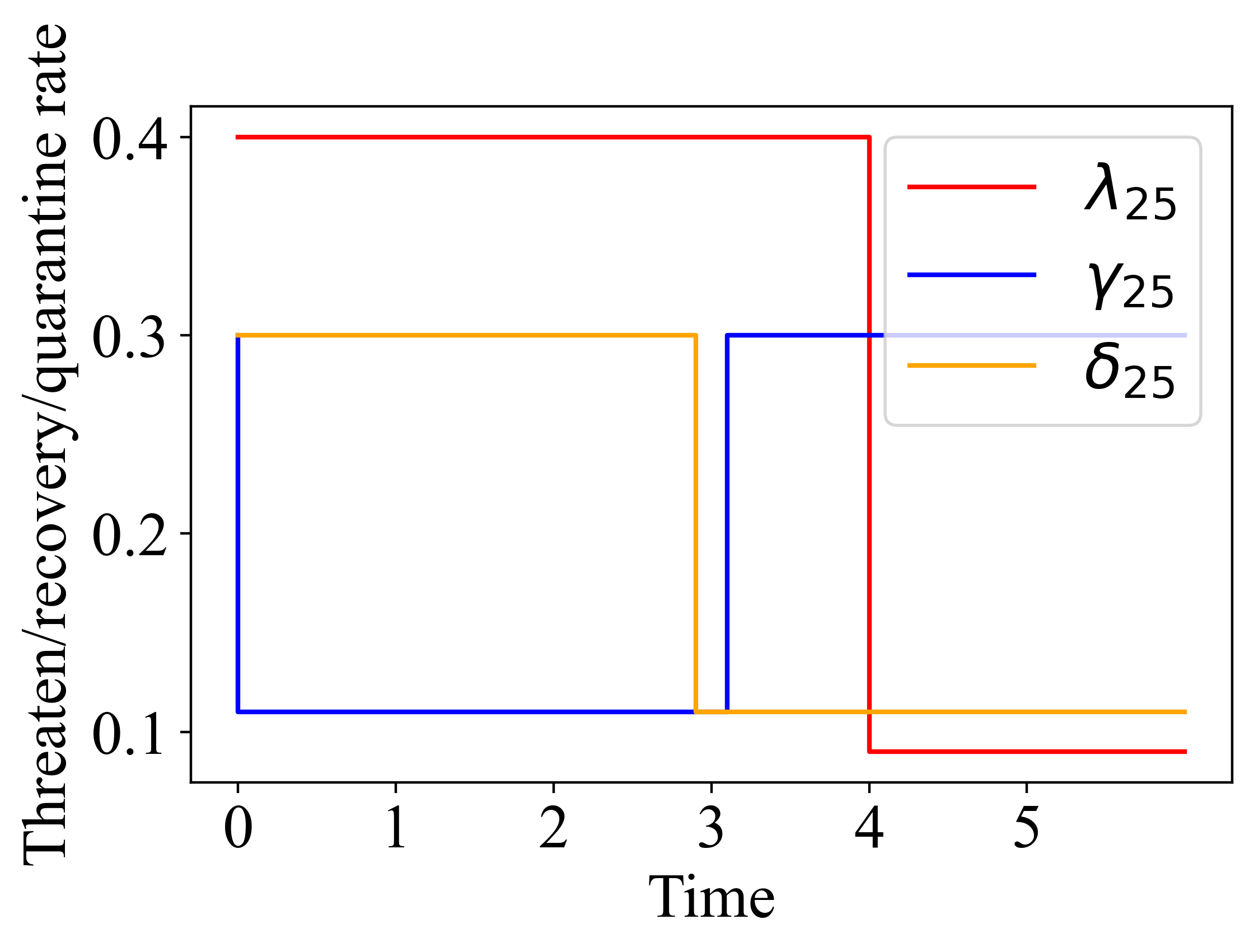}%
\label{fig_overallrepairexample3}}
\subfloat[\label{fig:c}]{\includegraphics[scale=0.3]{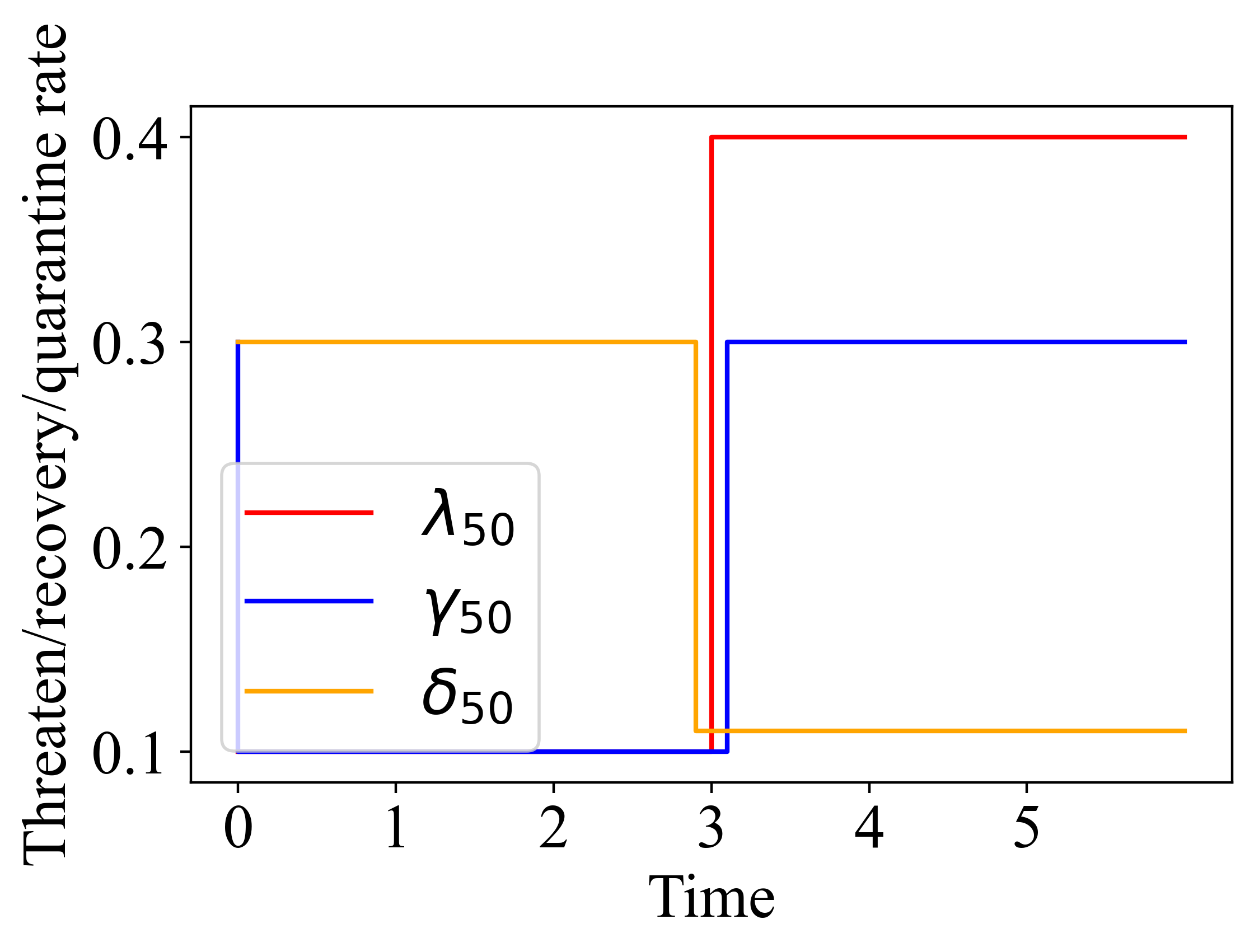}%
\label{fig_quarantinexample3}}
\\
\subfloat[\label{fig:b}]{\includegraphics[scale=0.3]{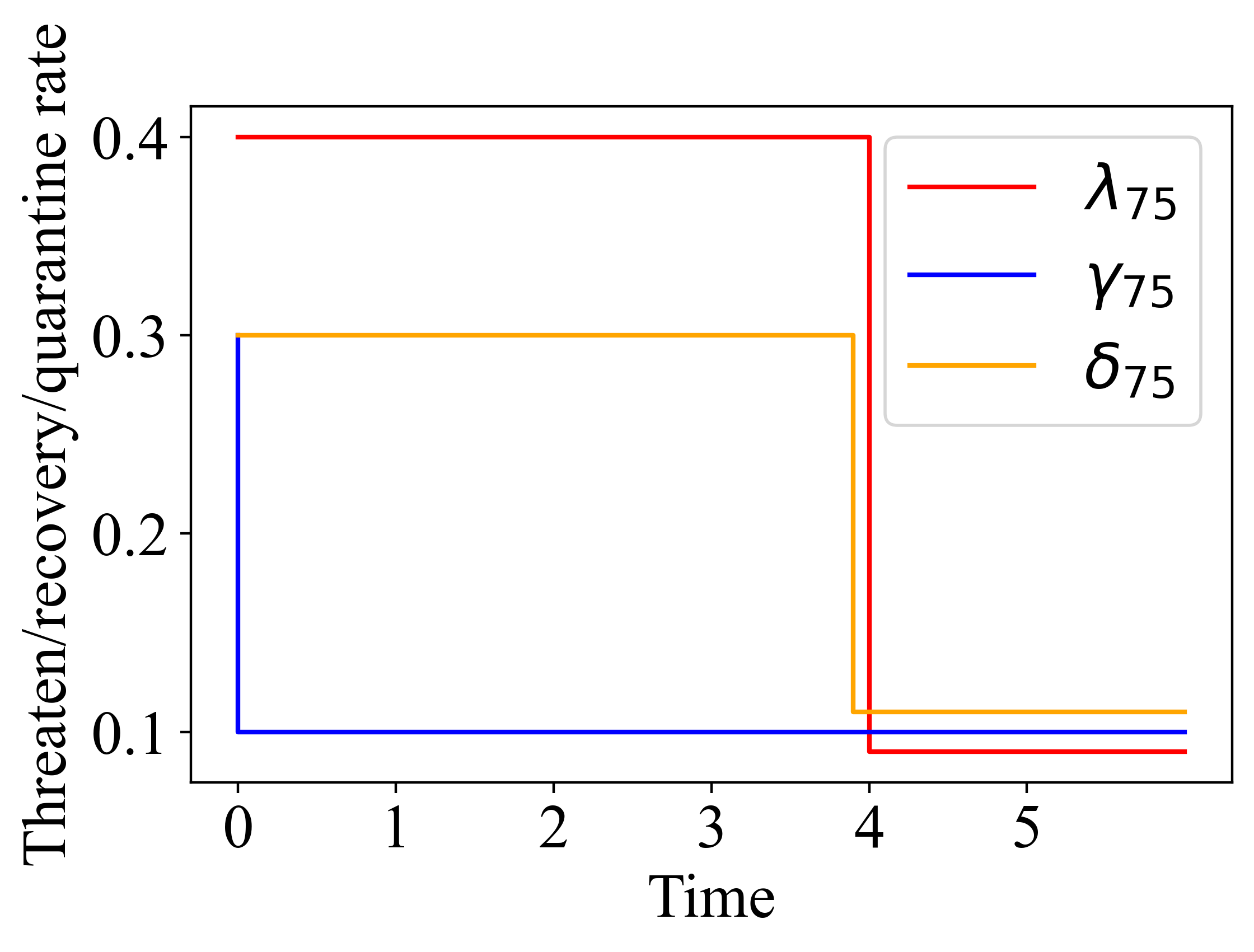}%
\label{fig_serviceultilityexample3}}
\subfloat[\label{fig:d}]{\includegraphics[scale=0.3]{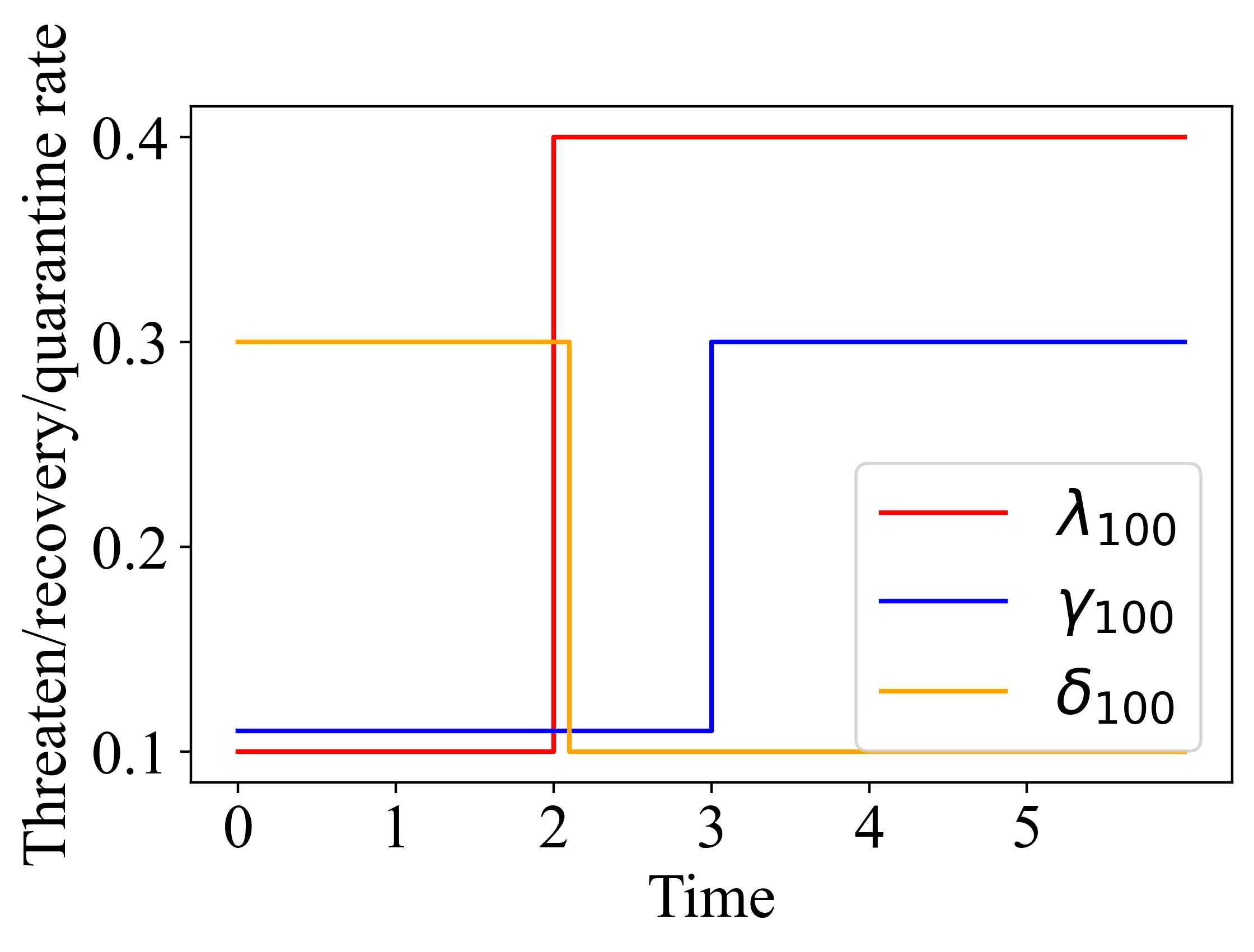}%
\label{fig_serviceultilityexample3}}

\caption{A sketch of the PPAC strategy in setting 1.}
\label{fig_ansofexample3}
\end{figure} 
\subsection{Numerical Results of Static Networks (Setting 1)}\label{static network}
Using the Pajek software \cite{de2018exploratory}, we stimulate a static network $G_{ST}$ with $N = 100$ nodes, the edge-rewiring probability is 0.1 and maintain network connectivity. The $A_{ST}(t)$ represents the adjacency matrix of $G_{ST}$ at time $t$ and $\bm{\hat{T}}$ represents the graph sequence of $G_{ST}$.\par
   Consider the APT repair game in static Networks:
   \begin{equation}
	\begin{aligned}
	\mathbb{M}=(G_{ST},\bm{a^{1*}_i},\bm{a^{2*}_i},\bm{b_i}^*,\bm{\phi},\bm{\varrho^1},\bm{\varrho^2},\bm{\alpha},\bm{\beta},\bm{\delta}^1,\bm{\delta}^4,\\
\bm{\lambda}^1,\bm{\lambda}^6,\bm{\gamma}^1,\bm{\gamma}^5,\bm{\hat{T}},
R,T,\bm{U_n},\bm{U},\bm{E}_0),\nonumber
	\end{aligned}
	\end{equation}
  where $G=G_{ST}$, $T = 12$, and $\bm{\hat{T}}=\{A_1=(0,2),A_1=(2,3),A_1=(3,4),A_1=(4,6)\}$. We set the following parameters under \textbf{Setting 1}:
 
  \begin{enumerate}
	\item The utility of the entire network $\bm{U}=1000$.
	\item Maintain minimum service effectiveness requirements for the mission $\bm{U_n} = 80\% \bm{U}$.
	\item Threaten rate list and recovery rate list are obtained by executing the TRG and RRG algorithms.
	\item The specific representation of each function is
        \begin{equation}
        \begin{aligned}
        \begin{aligned}
        &\boldsymbol{\phi}(\delta) =(\sqrt{\delta}, \ldots, \sqrt{\delta}),  \boldsymbol{\varrho^1}(\gamma) =\left(\sqrt{\gamma}, \ldots, \sqrt{\gamma}\right),
        \\
        &\boldsymbol{\varrho^2}(1-\gamma) =(\sqrt{1-\gamma}, \ldots, \sqrt{1-\gamma}).
        \end{aligned}\nonumber
    	\end{aligned}
    	\end{equation}
\end{enumerate}
\par

By executing the scheme, we get the potential defense strategy. Fig. \ref{fig_ansofexample3} (a)-(d) plot four threaten, quarantine, and recovery rate functions in the strategy. Fig. \ref{al} (a) represents the number of repair nodes at each moment in setting 1. We can see that the difference between the PPAC and ER methods for repair is not obvious under a fixed network, which is also in line with our initial assumptions. Because without considering subsequent topology changes, our method still requires decisions on quarantine, repair, and local repair based on effectiveness and cost. Fig. \ref{al} (b) represents the number of quarantined nodes at each moment in setting 1. The difference between PPAC and QAR in quarantined nodes is also insignificant. Still, the difference exists because our restrictions require that the number of nodes in the isolation zone can not exceed the threshold. Fig. \ref{al} (c) represents the service quality at each moment in setting 1. \par

In summary, the resource occupancy of PPAC, QAR, and ER are  13.43$\%$, 15.21$\%$, and 13.5$\%$, respectively. The defense resource utilization of PPAC rises 48.94$\%$, QAR rises 31.46$\%$, and ER rises 48.15$\%$. The service stability of each method is as follows: PPAC 100$\%$, QAR 42.85$\%$, and ER 85.71$\%$.  

\begin{figure*}[!t]
\centering
\subfloat[]{\includegraphics[width=2.2in]{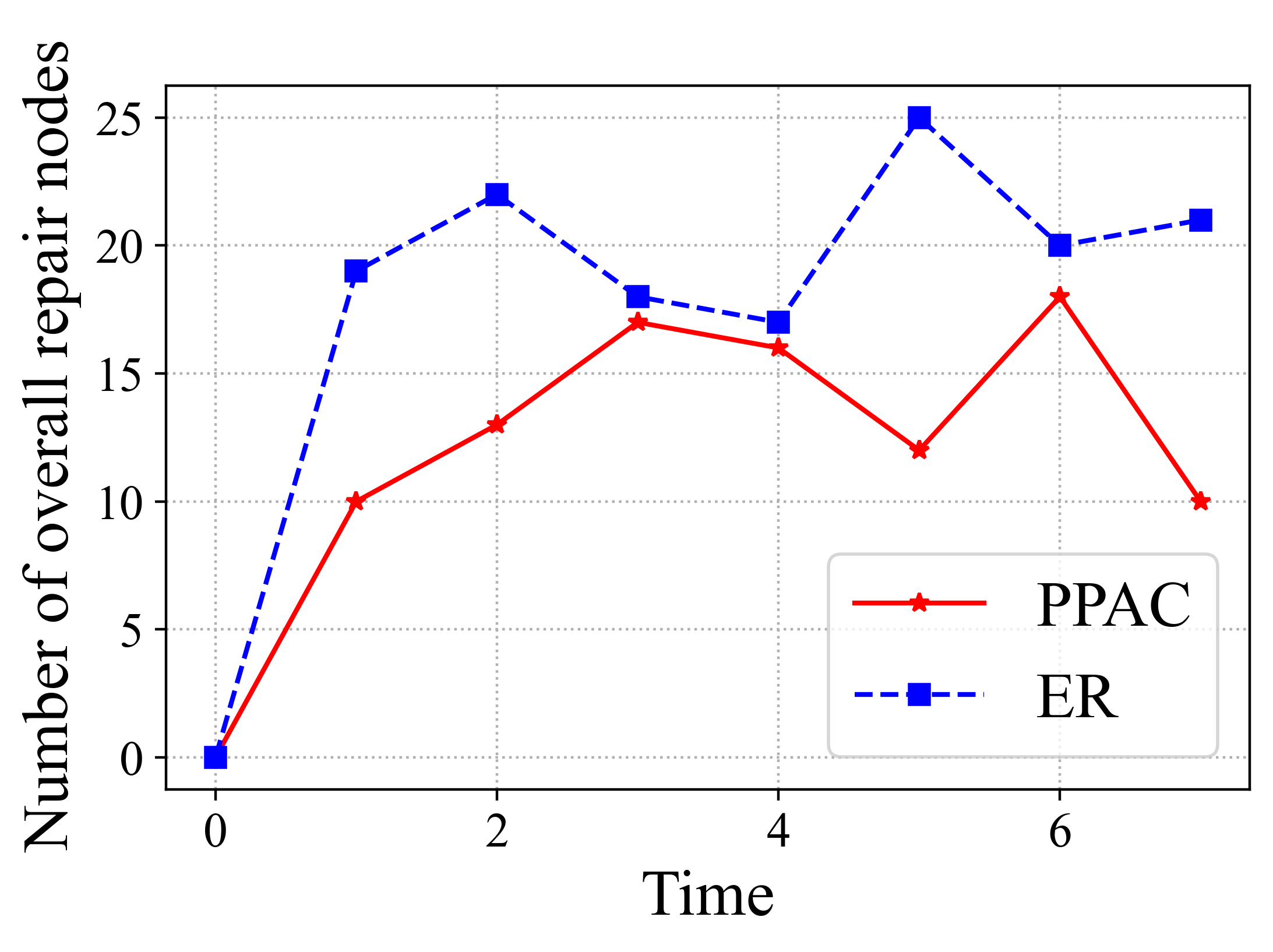}%
\label{fig_overallrepairexample1}}
\hfil
\subfloat[]{\includegraphics[width=2.2in]{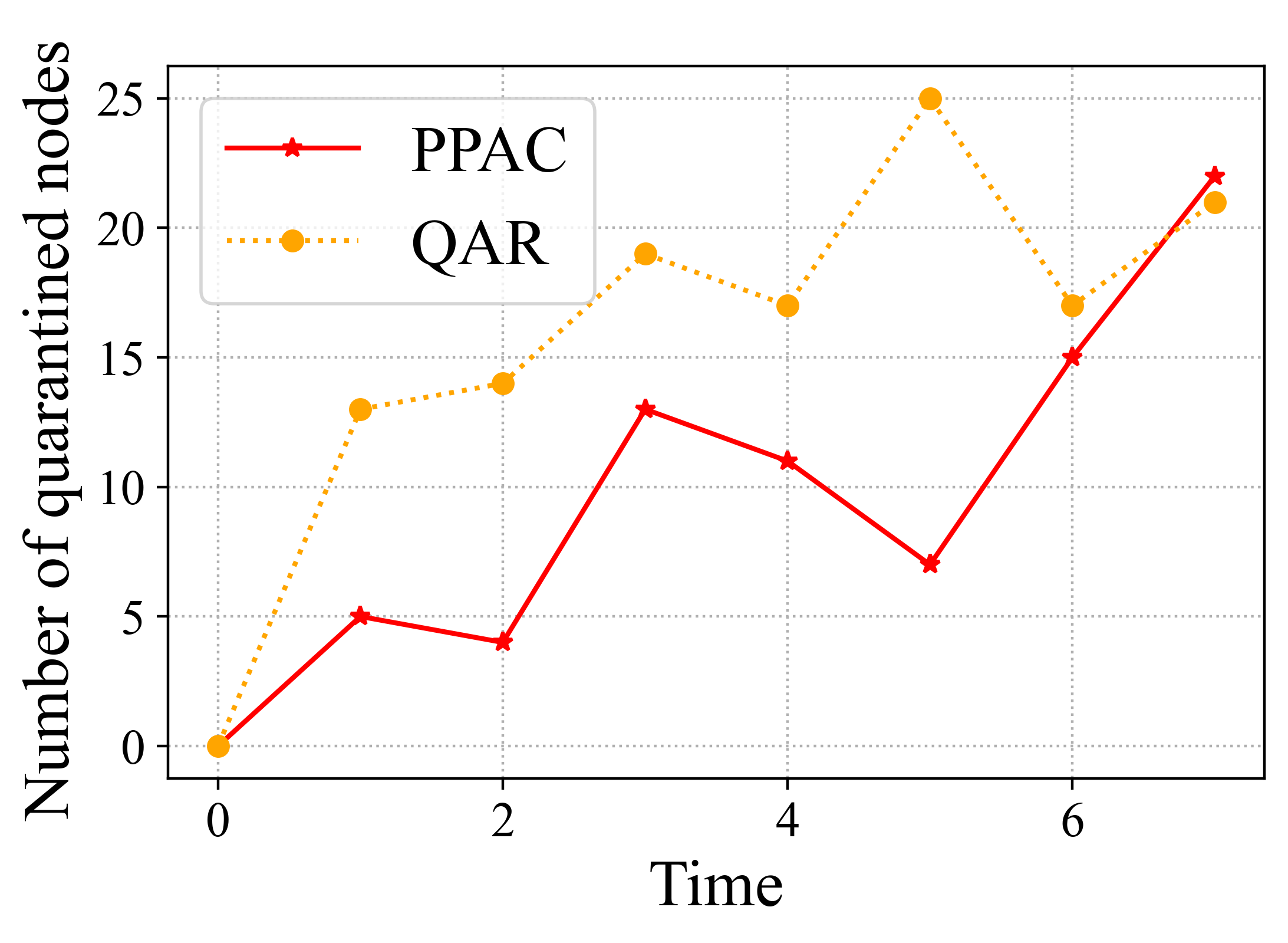}%
\label{fig_quarantinexample1}}
\hfil
\subfloat[]{\includegraphics[width=2.3in]{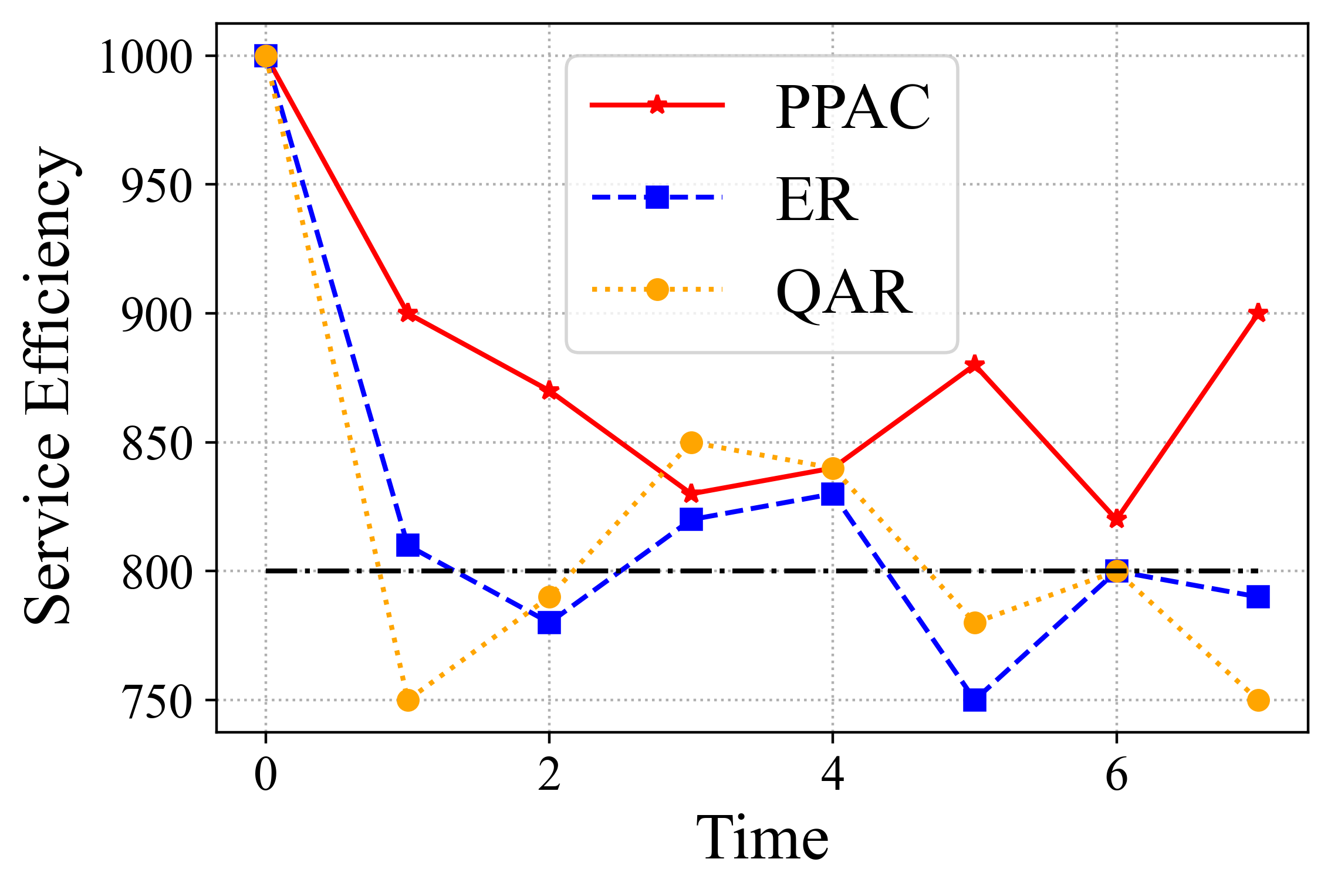}%
\label{fig_serviceultilityexample1}}
\hfil
\caption{(a) the comparison between our method and the ER by the number of nodes for overall maintenance at each moment in setting 2. (b) the comparison between our method and the QAR by the number of quarantined nodes at each moment in setting 2. (c) the comparison between our method and ER, QAR by the service quality at each time.}
\label{a2}
\end{figure*}

\begin{figure}[!h]
\centering
\subfloat[\label{fig:a}]{\includegraphics[scale=0.3]{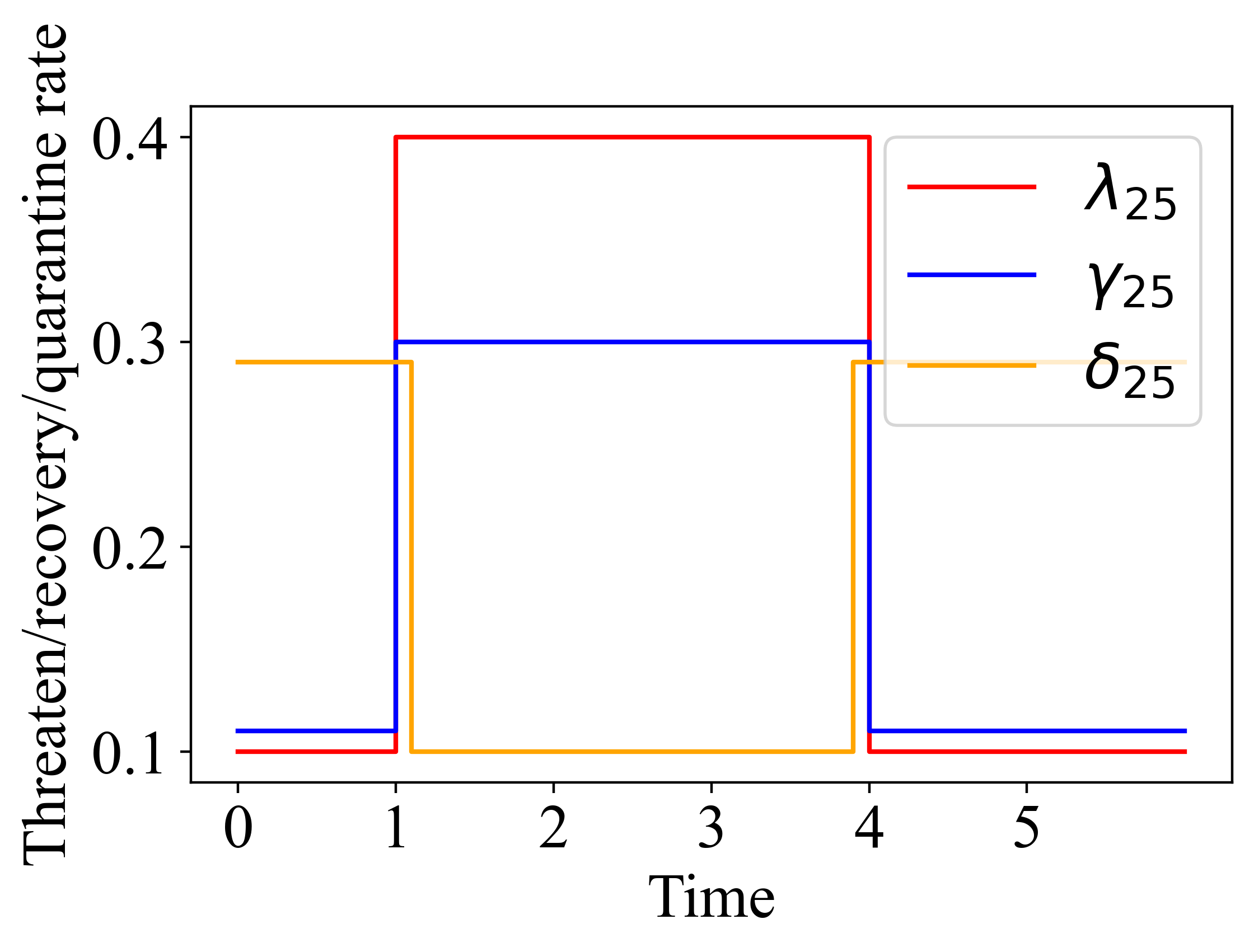}%
\label{fig_overallrepairexample3}}
\subfloat[\label{fig:c}]{\includegraphics[scale=0.3]{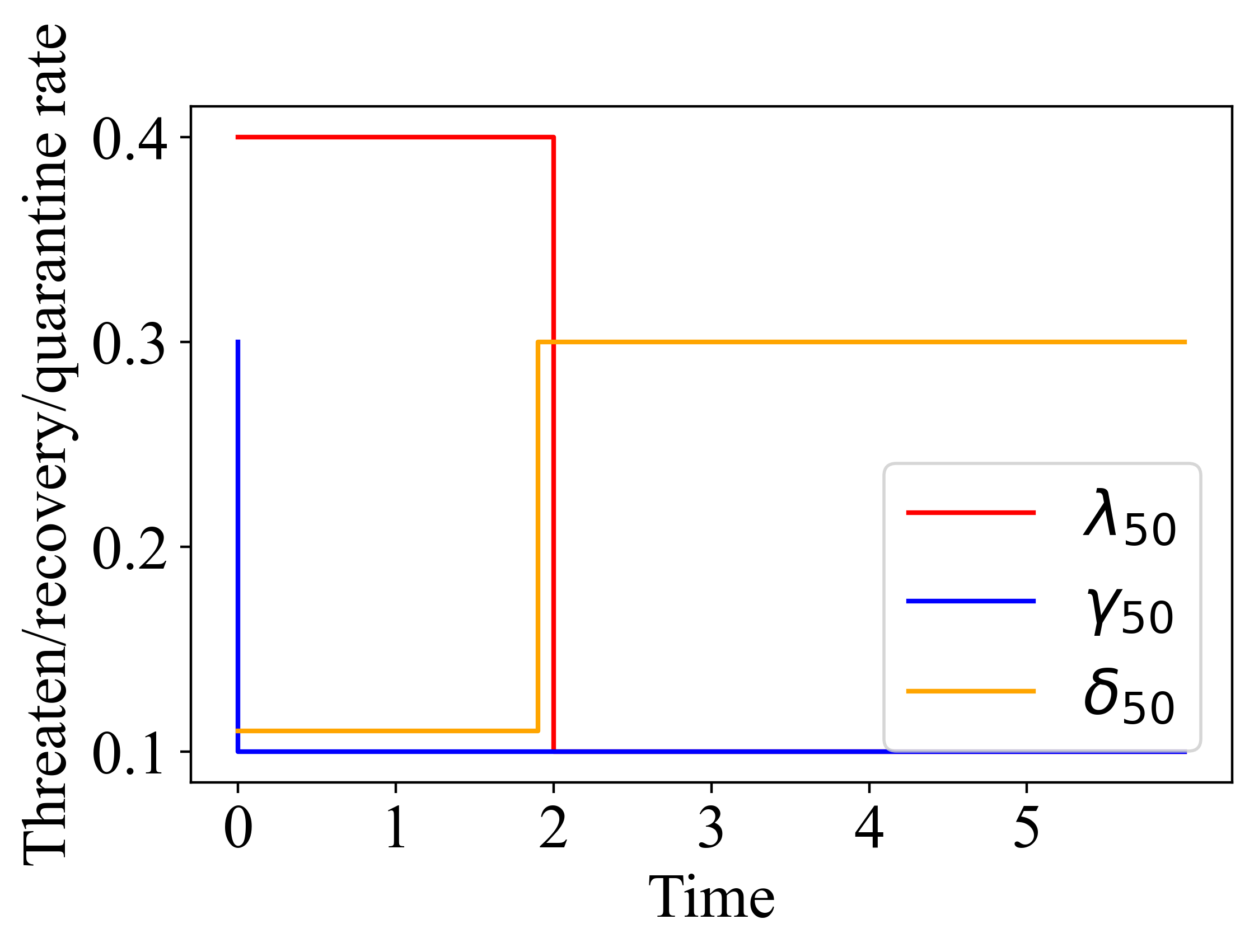}%
\label{fig_quarantinexample3}}
\\
\subfloat[\label{fig:b}]{\includegraphics[scale=0.3]{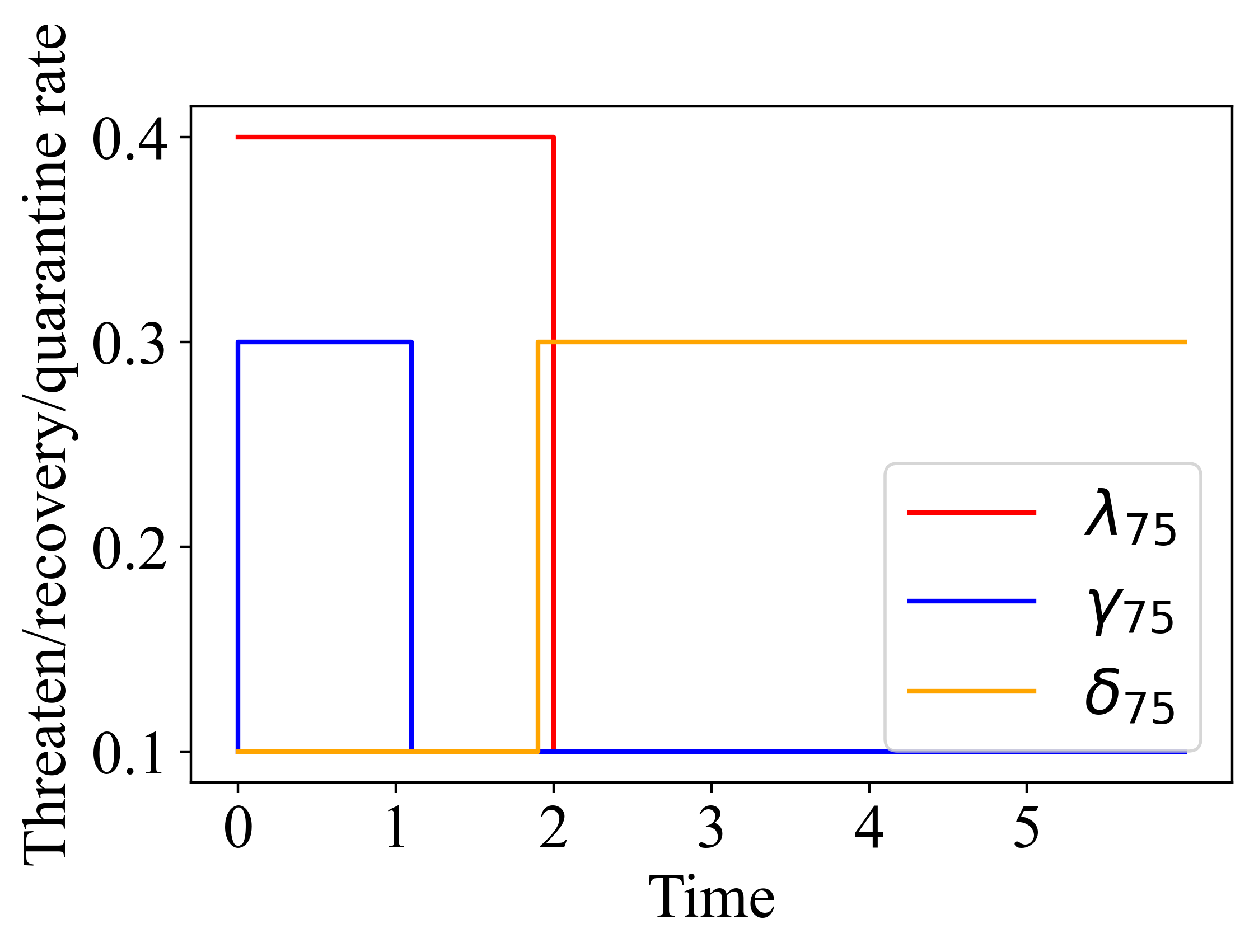}%
\label{fig_serviceultilityexample3}}
\subfloat[\label{fig:d}]{\includegraphics[scale=0.3]{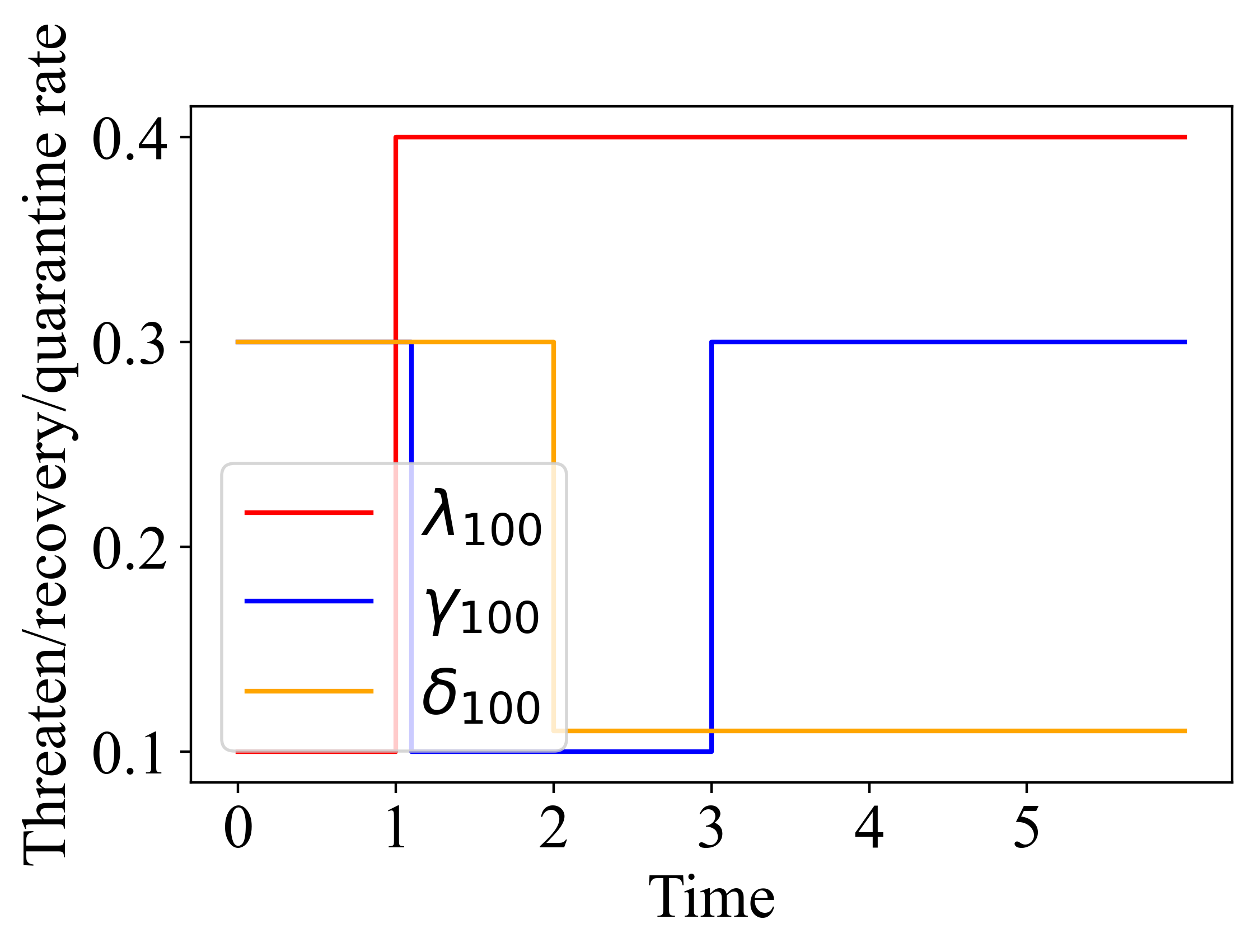}%
\label{fig_serviceultilityexample3}}

\caption{A sketch of the PPAC strategy in setting 2.}
\label{fig_ansofexample1}
\end{figure}

\subsection{Numerical Results of Periodic Time-varying Network (Setting~2)}\label{periodic network}
Using the Pajek software \cite{de2018exploratory}, we stimulate a periodic time-varying network $G_{PT}$ with $N = 100$ nodes, the edge-rewiring probability is 0.1 and maintain network connectivity. The $A_{PT}(t)$ represents the adjacency matrix of $G_{PT}$ at time $t$ and $\bm{\hat{T}}$ represents the graph sequence of $G_{PT}$.\par
   Consider the APT repair game in periodic time-varying Network:
   \begin{equation}
	\begin{aligned}
	\mathbb{M}=(G_{PT},\bm{a^{1*}_i},\bm{a^{2*}_i},\bm{b_i}^*,\bm{\phi},\bm{\varrho^1},\bm{\varrho^2},\bm{\alpha},\bm{\beta},\bm{\delta}^1,\bm{\delta}^4,\\
\bm{\lambda}^1,\bm{\lambda}^6,\bm{\gamma}^1,\bm{\gamma}^5,\bm{\hat{T}},
R,T,\bm{U_n},\bm{U},\bm{E}_0),\nonumber
	\end{aligned}
	\end{equation}
  where $G=G_{PT}$, $T = 6$, and
 $\bm{\hat{T}}=\{A_1=(0,2),A_2=(2,3),A_3=(3,4),A_4=(4,6)\}$. We set the following parameters under \textbf{Setting 2}:
 \begin{enumerate}
	\item The utility of the entire network $\bm{U}=1000$.
	\item Maintain minimum service effectiveness requirements for the mission $\bm{U_n} = 80\% \bm{U}$.
	\item Threaten rate list and recovery rate list T and R are obtained by executing the TRG and RRG algorithms.
	\item The specific representation of each function is
        \begin{equation}
        \begin{aligned}
        \begin{aligned}
        &\boldsymbol{\phi}(\delta) =(\sqrt{\delta}, \ldots, \sqrt{\delta}),  \boldsymbol{\varrho^1}(\gamma) =\left(\sqrt{\gamma}, \ldots, \sqrt{\gamma}\right),
        \\
        &\boldsymbol{\varrho^2}(1-\gamma) =(\sqrt{1-\gamma}, \ldots, \sqrt{1-\gamma}).
        \end{aligned}\nonumber
    	\end{aligned}
    	\end{equation}
\end{enumerate}
\par

By executing the scheme, we get the potential defense strategy. Fig. \ref{fig_ansofexample1} (a)-(d) plots four threat, quarantine, and recovery rate functions in the strategy. From the figure, we can see that the importance of our approach to the node's service effectiveness and the overall maintenance of resources results in less maintenance of the attacked nodes in the early stages of the whole task. Therefore, it leads to a later surge in the number of nodes to maintain effectiveness and mitigate the previous impact. Fig. \ref{a2} (a) shows the number of overall maintenance for the node at each moment in setting 2. We can see that the overall curve of the number of repairs for PPAC at each moment is significantly lower than ER, which shows that PPAC achieves accurate control under periodic time-varying networks. Fig. \ref{a2} (b) represents the number of quarantined nodes at each moment in setting 2. The overall curve of the number of quarantined nodes for PPAC at each moment is significantly lower than QAR.

\par

In summary, the resource occupancy of PPAC, QAR, and ER are 9.61$\%$, 14.5$\%$, and 15.14$\%$, respectively. The defense resource utilization of PPAC rises 61.85$\%$, QAR rises 37.93$\%$, and ER rises 32.08$\%$. The service stability of each method is as follows: PPAC 100$\%$, QAR 41.67$\%$, and ER 66.67$\%$.
\subsection{Numerical Results of Time-varying Network (Setting 3)}\label{time-varying network}
Using the Pajek software \cite{de2018exploratory}, we stimulate a time-varying network $G_{TV}$ with $N = 100$ nodes, the edge-rewiring probability is 0.1 and maintain network connectivity. The $A_{TV}(t)$ represents the adjacency matrix of $G_{TV}$ at time $t$ and $\bm{\hat{T}}$ represents the graph sequence of $G_{TV}$.\par

   Consider the APT repair game in time-varying Network:
   \begin{equation}
	\begin{aligned}
	\mathbb{M}=(G_{TV},\bm{a^{1*}_i},\bm{a^{2*}_i},\bm{b_i}^*,\bm{\phi},\bm{\varrho^1},\bm{\varrho^2},\bm{\alpha},\bm{\beta},\bm{\delta}^1,\bm{\delta}^4,\\
\bm{\lambda}^1,\bm{\lambda}^6,\bm{\gamma}^1,\bm{\gamma}^5,\bm{\hat{T}},
R,T,\bm{U_n},\bm{U},\bm{E}_0),\nonumber
	\end{aligned}
	\end{equation}
  where $G=G_{TV}$, $T = 12$, and
 $\bm{\hat{T}}=\{A_1=(0,2),A_2=(2,5),A_3=(5,6),A_4=(6,9),A_5=(9,11),A_6=(11,12)\}$. 
We conduct a statistical analysis based on the structural changes of the Xi'an Jiaotong University AWD platform (Attack-Defense War platform) and the data from attack-defense exercises on the platform, to provide real-world settings for time-varying networks. We set the following parameters under \textbf{Setting 3}:
 \begin{enumerate}
	\item The utility of the entire network $\bm{U}=1000$.
	\item Maintain minimum service effectiveness requirements for the mission $\bm{U_n} = 80\% \bm{U}$.
	\item Threaten rate list and recovery rate list T and R are obtained by executing the TRG,RRG algorithms.
	\item The specific representation of each function is
        \begin{equation}
        \begin{aligned}
        \begin{aligned}
        &\boldsymbol{\phi}(\delta) =(\sqrt{\delta}, \ldots, \sqrt{\delta}),  \boldsymbol{\varrho^1}(\gamma) =\left(\sqrt{\gamma}, \ldots, \sqrt{\gamma}\right),
        \\
        &\boldsymbol{\varrho^2}(1-\gamma) =(\sqrt{1-\gamma}, \ldots, \sqrt{1-\gamma}).
        \end{aligned}\nonumber
    	\end{aligned}
    	\end{equation}
\end{enumerate}
\begin{figure*}[!t]
\centering
\vspace{-0.5cm}
\subfloat[]{\includegraphics[width=2.2in]{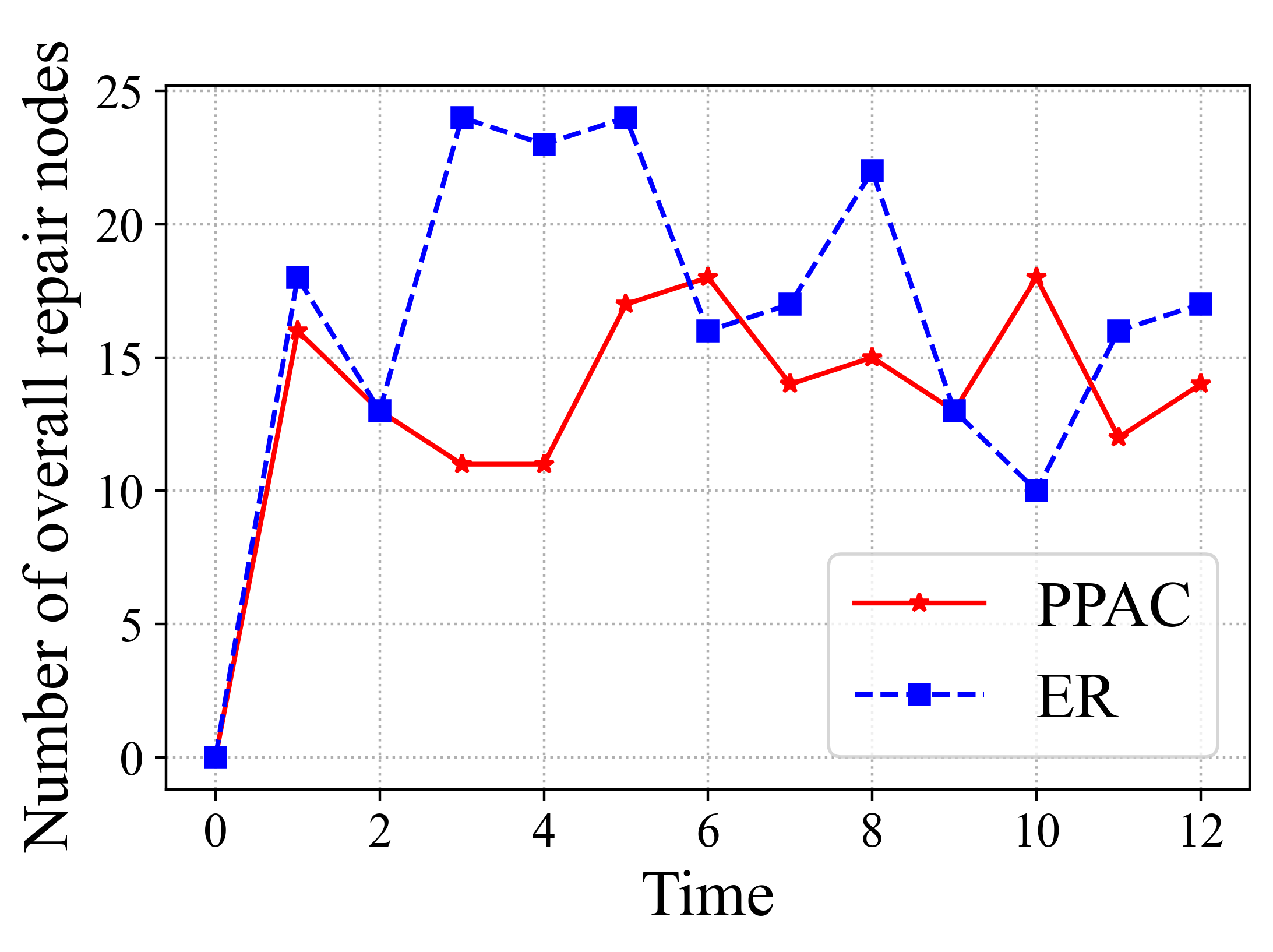}%
\label{fig_overallrepairexample2}}
\hfil
\subfloat[]{\includegraphics[width=2.2in]{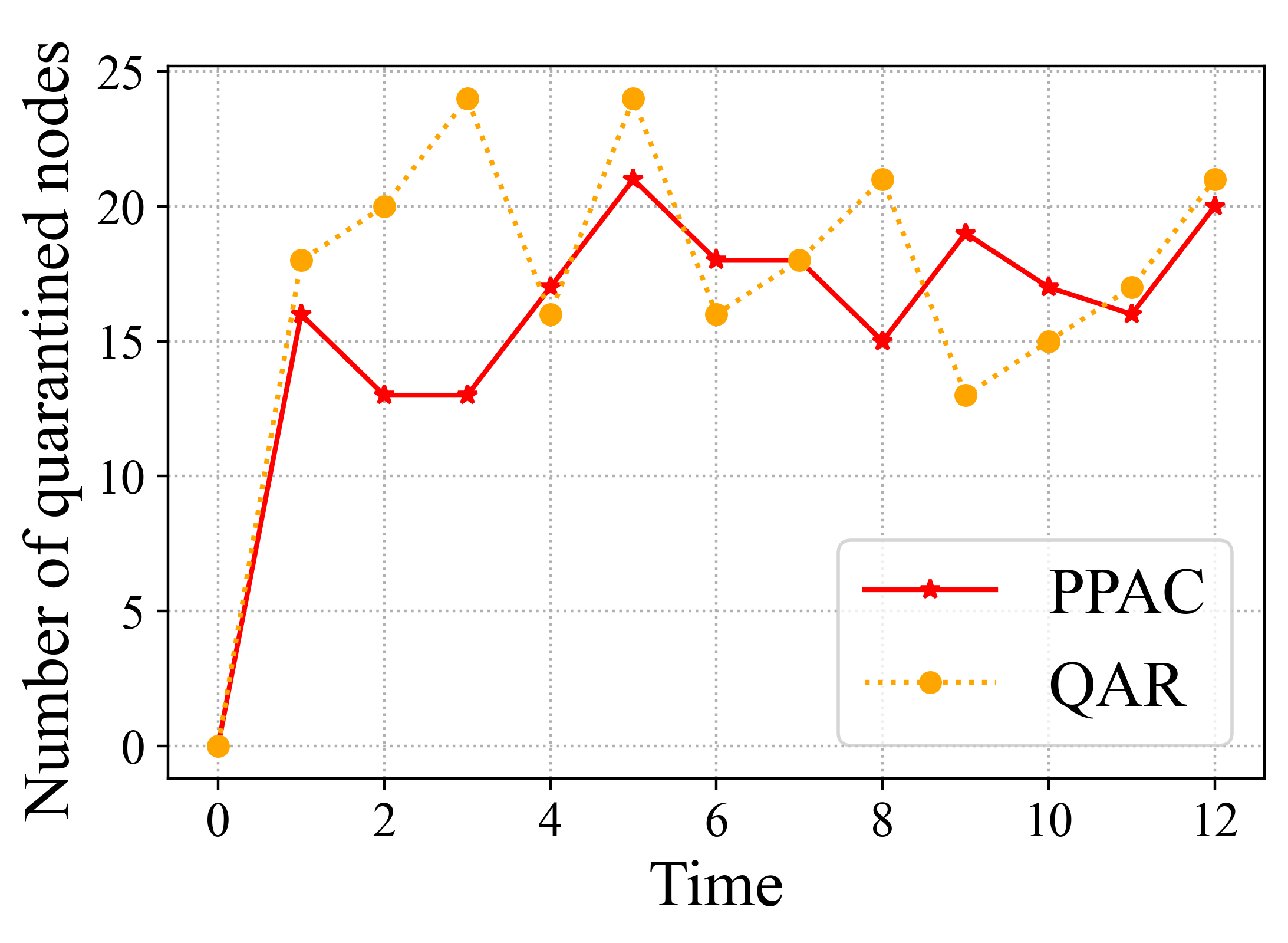}%
\label{fig_quarantinexample2}}
\hfil
\subfloat[]{\includegraphics[width=2.3in]{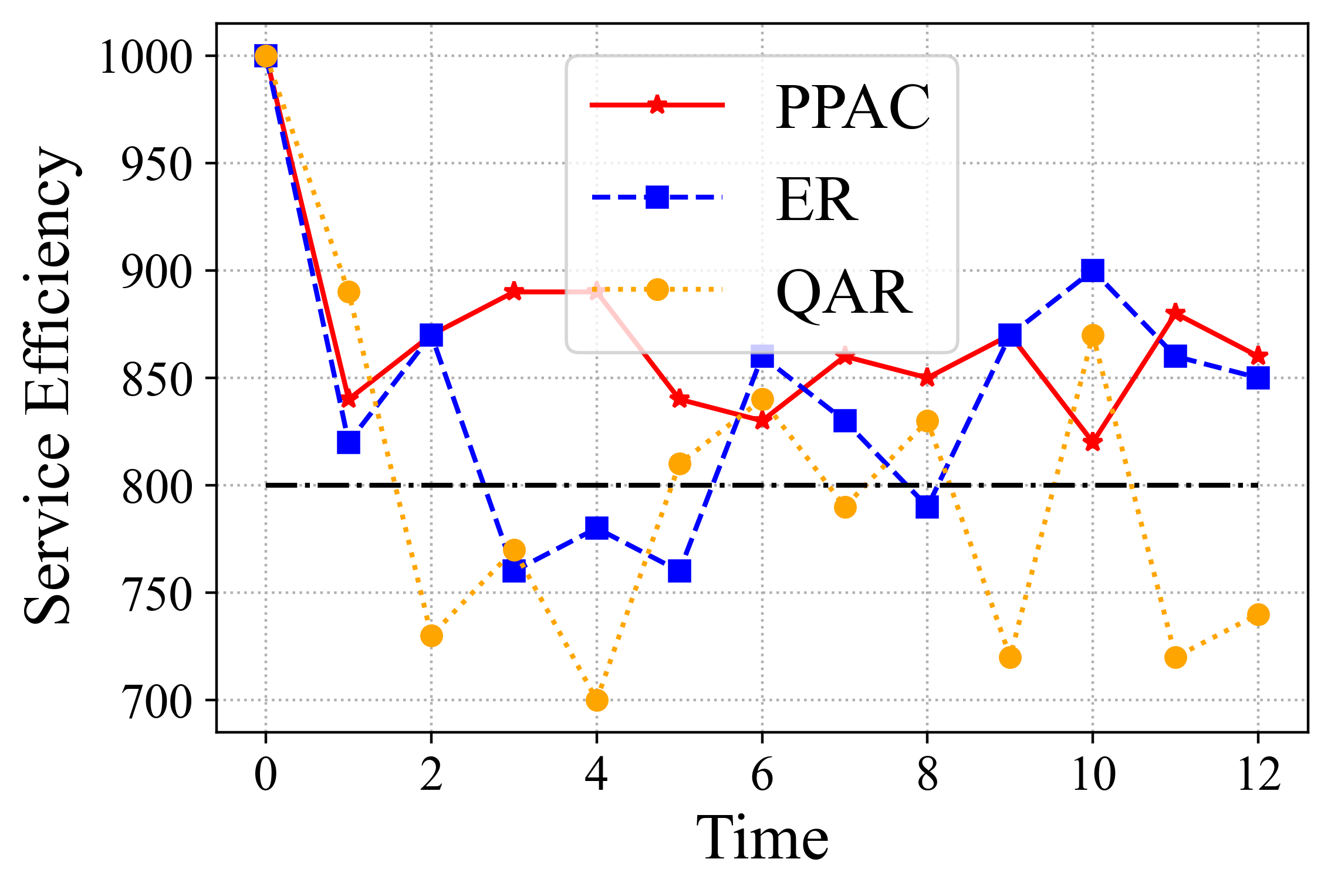}%
\label{fig_serviceultilityexample2}}
\hfil
\caption{(a) the comparison between our method and the ER by the number of nodes for overall maintenance at each moment in setting 3. (b) the comparison between our method and the QAR by the number of quarantined nodes at each moment in setting 3.(c) the comparison between our method and ER, QAR by the service quality at each time.}
\label{a3}
\end{figure*}
\begin{figure}[h]
\centering
\subfloat[\label{fig:a}]{\includegraphics[scale=0.3]{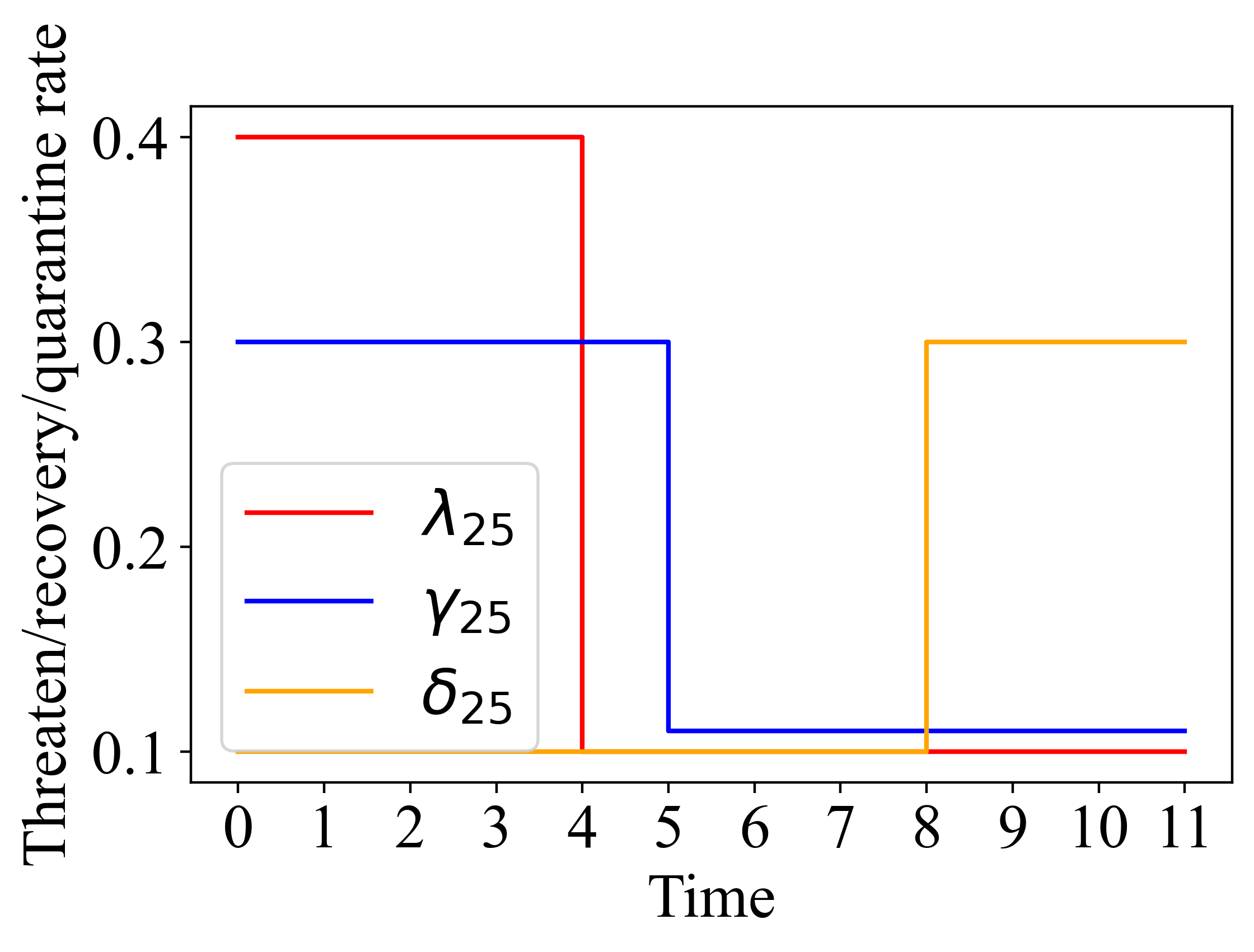}%
\label{fig_overallrepairexample3}}
\subfloat[\label{fig:c}]{\includegraphics[scale=0.3]{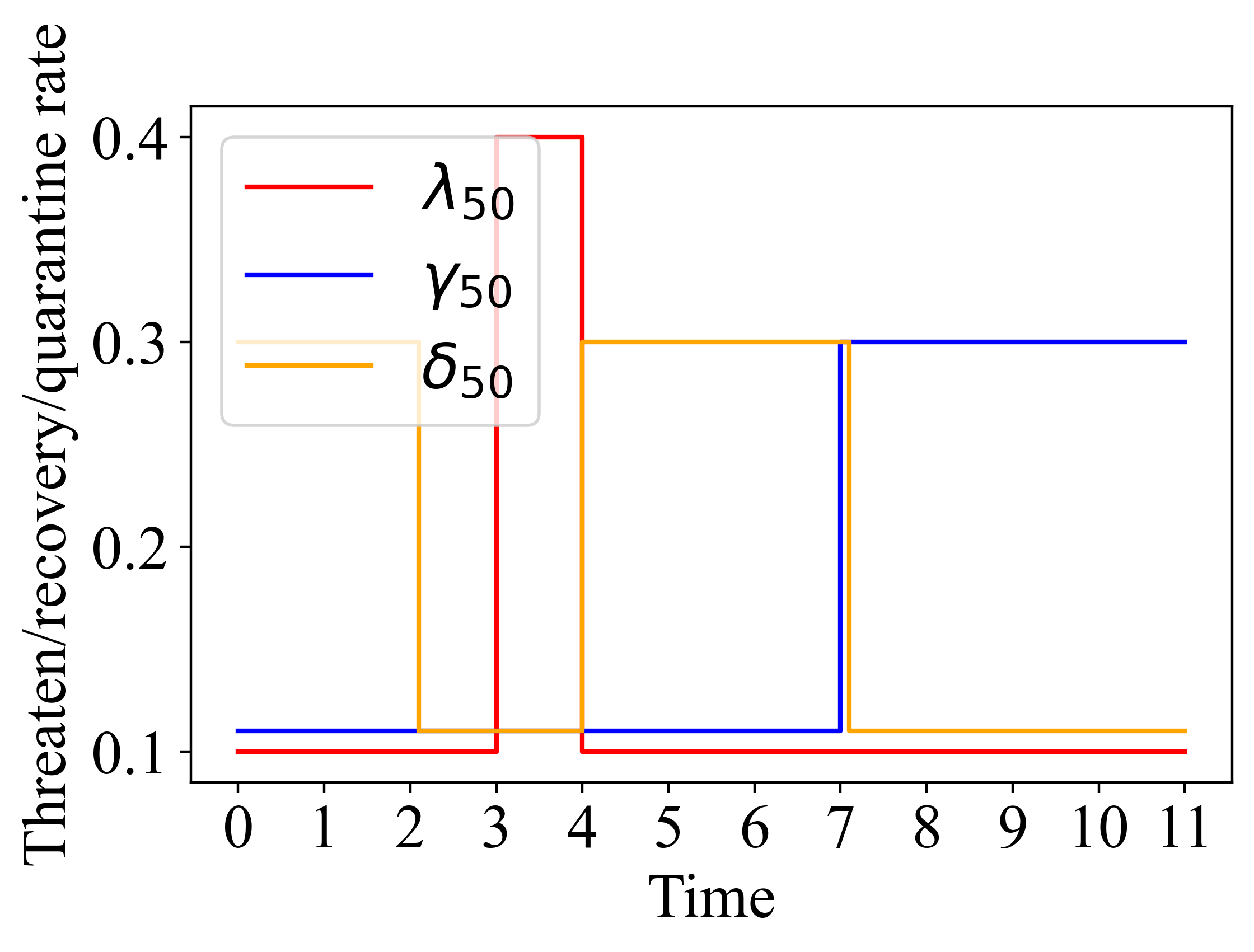}%
\label{fig_quarantinexample3}}
\\
\subfloat[\label{fig:b}]{\includegraphics[scale=0.3]{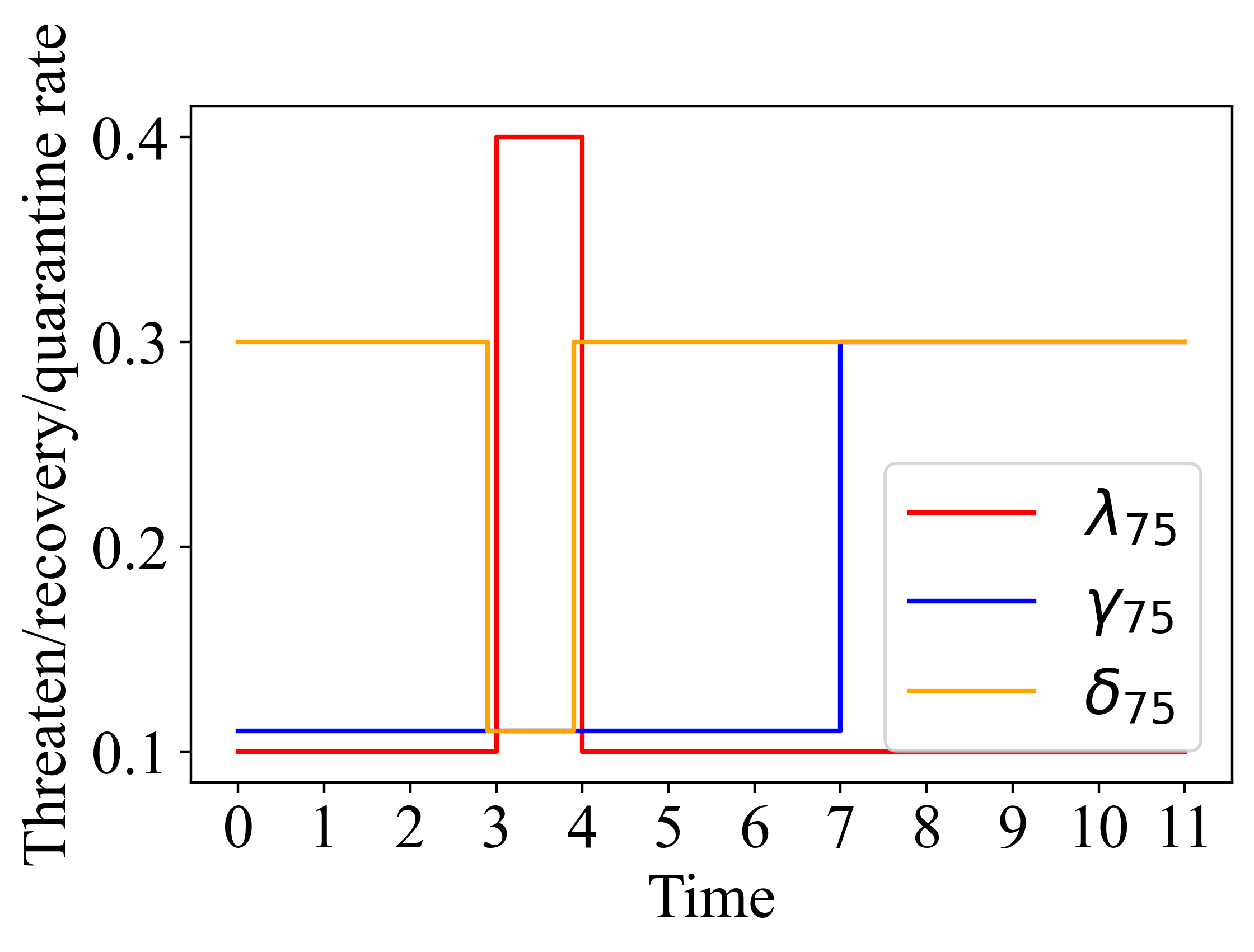}%
\label{fig_serviceultilityexample3}}
\subfloat[\label{fig:d}]{\includegraphics[scale=0.3]{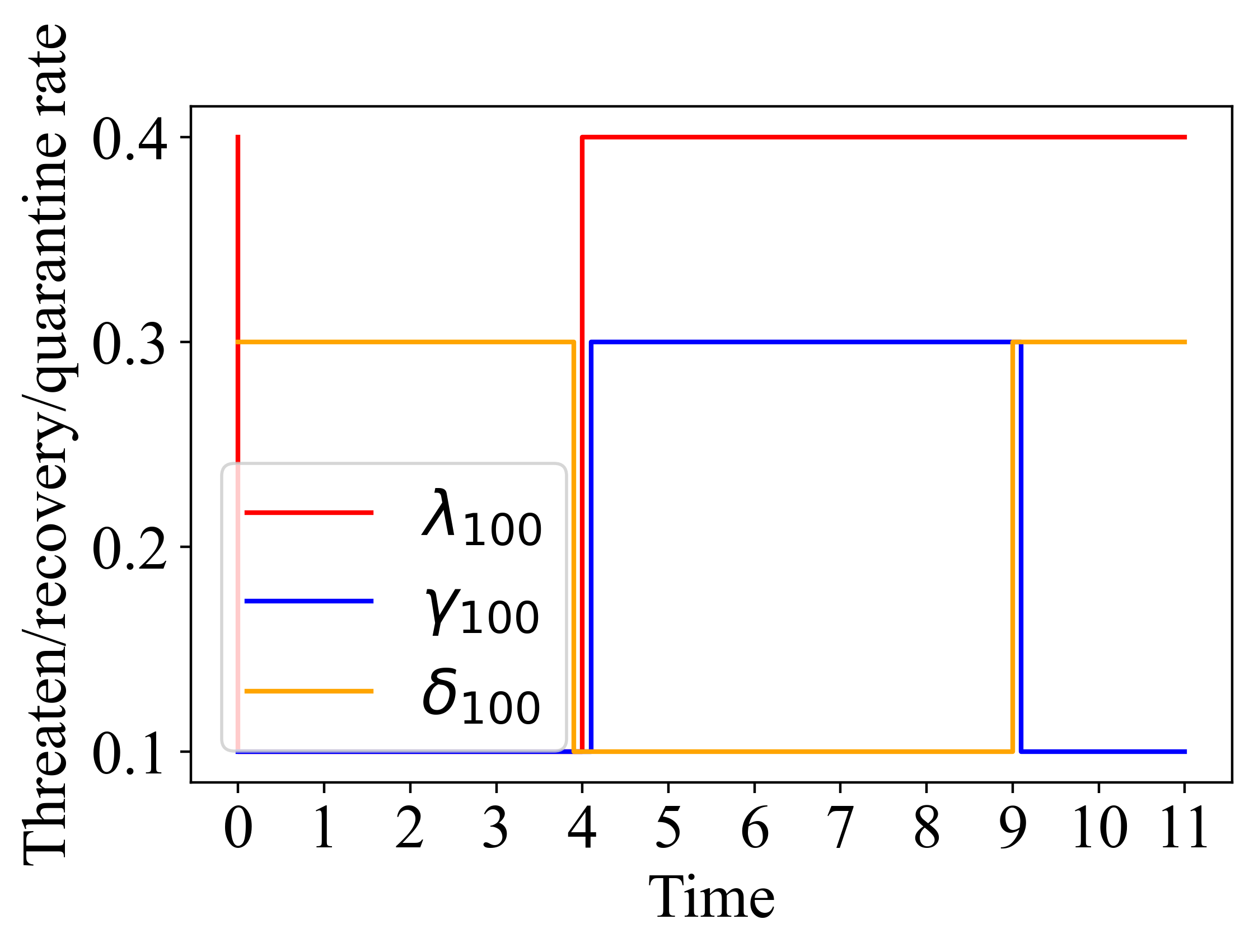}%
\label{fig_serviceultilityexample3}}
\caption{A sketch of the PPAC strategy in setting 3.}
\label{fig_ansofexample2}
\end{figure}
By executing the scheme, we get the potential defense strategy. Fig. \ref{fig_ansofexample2} (a)-(d) plots threat, quarantine, and recovery rate functions in the strategy. Fig. \ref{a3} (a) represents the number of overall maintenance for the node at each moment in setting 3. We can see that the overall curve of the number of repairs for PPAC at each moment is significantly lower than ER, which shows that our solution achieves accurate control under time-varying networks. Fig. \ref{a3} (b) represents the number of quarantined nodes at each moment in setting 3. The overall curve of the number of quarantined nodes for PPAC at each moment is significantly lower than QAR. PPAC maintains a stable and high-quality service performance of the network and achieves the victim status of the node. Fig. \ref{a3} (c) represents the service quality at each moment in setting 3. We can see that all the methods except PPAC have a performance lower than the required performance at some time. \par

In summary, the resource occupancy of PPAC, QAR, and ER are 11.40$\%$, 10.48$\%$, and 11.79$\%$, respectively. The defense resource utilization of PPAC rises 75.50$\%$, QAR rises 36.56$\%$, and ER rises 44.14$\%$. The service stability of each method is as follows: PPAC 100$\%$, QAR 58.3$\%$, and ER 66.67$\%$. 
\subsection{Key Insights}
Our simulation is based on the rules of AWD (Attack with Defence) competitions \cite{CTFtime, CTFwiki}, and the strategies we derive can be directly applied during the exercise. For example,  \textbf{Setting 3} involves a dynamic attack and defense battle lasting 12 hours, and the strategies obtained from our method can be directly used by the defenders. In other words, the $\mathbf{\Lambda}(t),\mathbf{\Delta}(t),\mathbf{\Gamma}(t)$ in strategy $\bm{w}(t)$ represent the recommended probabilities for performing specific actions on each node.
\par
 In the real world, defenders can simulate such exercises by placing their own organizations within a set of rules tailored to their needs. By simulating these exercises and obtaining strategies during the exercise, defenders can identify weaknesses in their organization, make adjustments, and receive tactical training to better respond to attacks.
\section{SUMMARY and Future Work}
This paper investigates a new problem of defending against APT attacks under time-varying networks. We model the interactions among attacks and defenders in an APT process as a dynamic APT repair game and derive its near-optimal solution by formulating the optimization problem as a PPAC problem. We derive a reliable solving mechanism that can search for a timely defense strategy with cost savings and efficiency. Specifically, our defense strategy assigns discriminatory weights for each node to reveal critical nodes and the attacking intent of the adversary. Experimental results validate the effectiveness of defense performance under a moderately sized and time-varying network. \par
For future work, it is interesting to analyze the influence of variation of nodes in the heterogeneous networks scenario where attackers do not know the participation strategies of different types of nodes or in the asymmetric information scenario where the node type and function are the secret information of defenders.

%

\bibliographystyle{IEEEtran}
\bibliography{APT.bib}{}


\IEEEaftertitletext{\vspace{-2.5in}}
\begin{IEEEbiography}[{\includegraphics[width=1in,height=1.25in,clip,keepaspectratio]{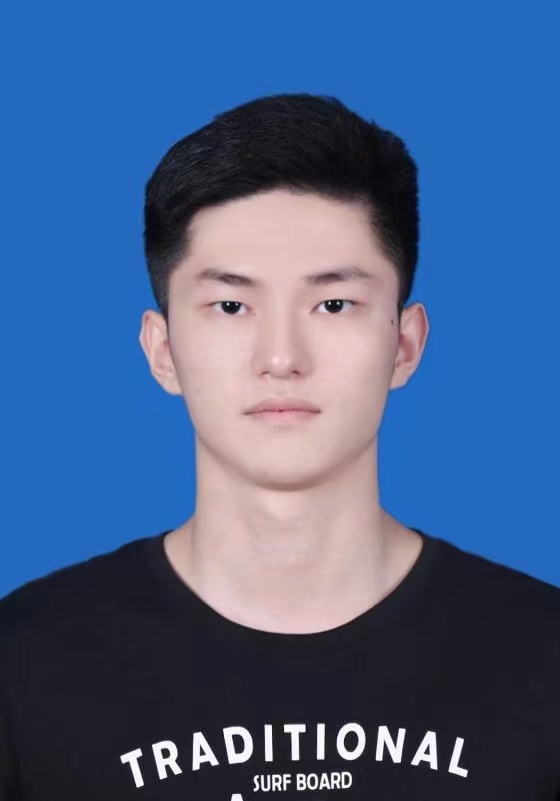}}]{Zixuan Wang}
received the B.E. degree from School of Materials Science and Engineering, Xi'an University of Architecture and Technology, Xi'an, China, in 2020. He is currently pursuing his M.S. degree in the School of Cyberspace Security, Xi'an Jiaotong University. His research interests include game theory, software and network security, and blockchain.\end{IEEEbiography}
\IEEEaftertitletext{\vspace{-2.5in}}
\begin{IEEEbiography}[{\includegraphics[width=1in,height=1.25in,clip,keepaspectratio]{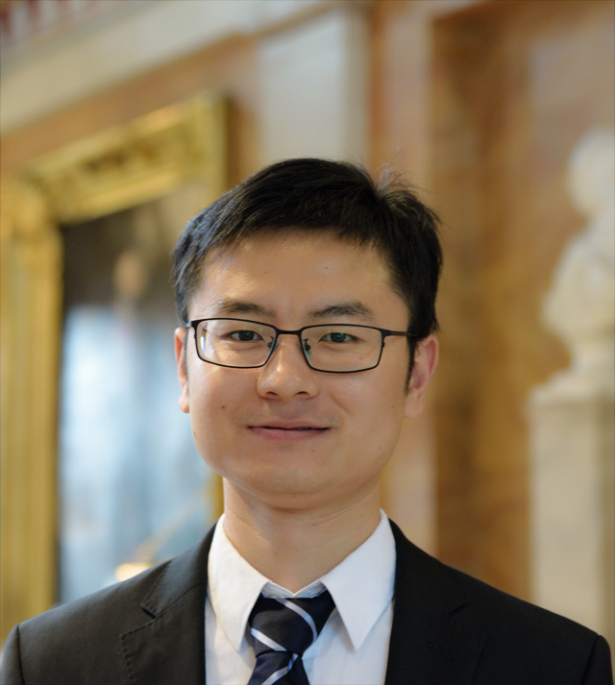}}]{Jiliang Li}
received the Dr. rer. nat. degree in computer science from the University of  G\"{o}ttingen, Germany. He is currently a Researcher Professor and PhD Supervisor with the School of Cyber Science and Engineering, Xi'an Jiaotong University, Xi'an, China. His research interests include information security, cryptography, blockchain and IoT security.
\end{IEEEbiography}
\IEEEaftertitletext{\vspace{-2.5in}}
\begin{IEEEbiography}[{\includegraphics[width=1in,height=1.25in,clip,keepaspectratio]{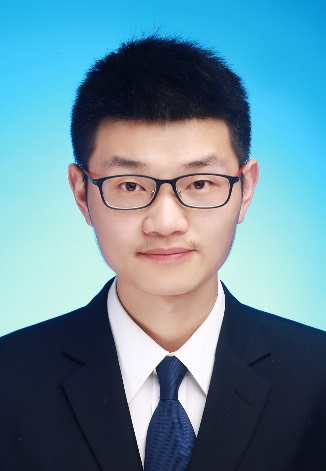 }}]{Yuntao Wang}
received the Ph.D degree in Cyberspace Security from Xi'an Jiaotong University, Xi'an, China, in 2022, where he is currently an Assistant Professor with the School of Cyber Science and Engineering. His research interests include security and privacy in intelligent IoT, network games, and blockchain.
\end{IEEEbiography}
\IEEEaftertitletext{\vspace{-2.5in}}
\begin{IEEEbiography}[{\includegraphics[width=1in,height=1.25in,clip,keepaspectratio]{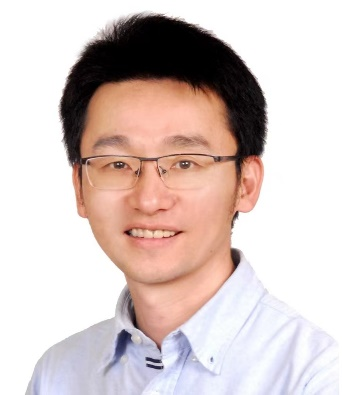}}]{Zhou Su}
has published technical papers, including top journals and top conferences, such as {\scshape IEEE Journal on Selected Areas in Communications}, {\scshape IEEE Transactions on Information Forensics and Security}, {\scshape IEEE Transactions on Dependable and Secure Computing}, {\scshape IEEE Transactions on Mobile Computing}, {\scshape IEEE/ACM Transactions on Networking}, and {\scshape INFOCOM}. 
Dr. Su received the Best Paper Award of International Conference IEEE ICC2020, IEEE BigdataSE2019, and IEEE CyberSciTech2017. He is an Associate Editor of {\scshape IEEE Internet of Things Journal}, {\scshape IEEE Open Journal of the Computer Society}, and {\scshape IET Communications}.
\end{IEEEbiography}
\IEEEaftertitletext{\vspace{-2.5in}}
\begin{IEEEbiography}[{\includegraphics[width=1in,height=1.25in,clip,keepaspectratio]{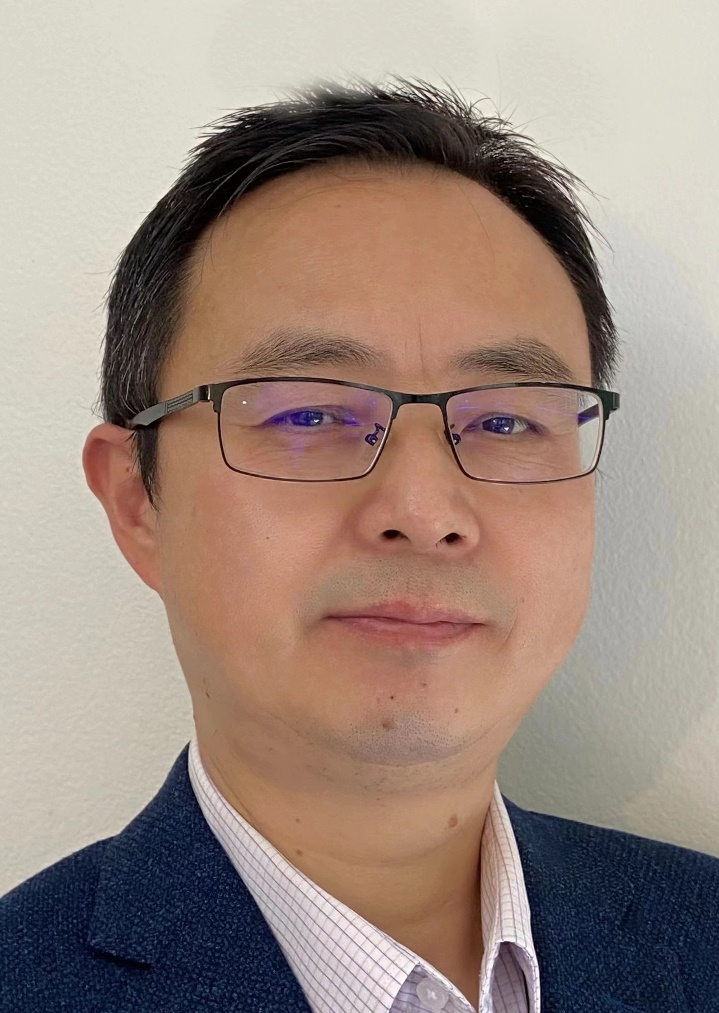}}]{Shui Yu (IEEE F’23)}
obtained his PhD from Deakin University, Australia, in 2004. He is a Professor of School of Computer Science, Deputy Chair of University Research Committee, University of Technology Sydney, Australia. His research interest includes Cybersecurity, Network Science, Big Data, and Mathematical Modelling. He has published five monographs and edited two books, more than 500 technical papers at different venues, such as IEEE TDSC, TPDS, TC, TIFS, TMC, TKDE, TETC, ToN, and INFOCOM. His current h-index is 71. Professor Yu promoted the research field of networking for big data since 2013, and his research outputs have been widely adopted by industrial systems, such as Amazon cloud security. He is currently serving the editorial boards of IEEE Communications Surveys and Tutorials (Area Editor) and IEEE Internet of Things Journal (Editor). He served as a Distinguished Lecturer of IEEE Communications Society (2018-2021). He is a Distinguished Visitor of IEEE Computer Society,  and an elected member of Board of Governors of IEEE VTS and ComSoc, respectively. He is a member of ACM and AAAS, and a Fellow of IEEE.
\end{IEEEbiography}
\IEEEaftertitletext{\vspace{-2.5in}}
\begin{IEEEbiography}[{\includegraphics[width=1in,height=1.25in,clip,keepaspectratio]{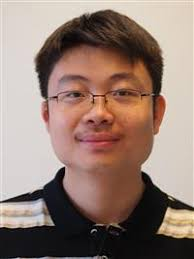}}]{Weizhi Meng}
 is currently an Associate Professor in the Department of Applied Mathematics and Computer Science, Technical University of Denmark (DTU), Denmark. He obtained his Ph.D. degree in Computer Science from the City University of Hong Kong, Hong Kong. Prior to joining DTU, he worked as Research Scientist in Institute for Infocomm Research, A*Star, Singapore. He won the Outstanding Academic Performance Award during his doctoral study, and is a recipient of the Hong Kong Institution of Engineers (HKIE) Outstanding Paper Award for Young Engineers/Researchers in both 2014 and 2017. He also received the IEEE ComSoc Best Young Researcher Award for Europe, Middle East, \& Africa Region (EMEA) in 2020. His primary research interests are cyber security and intelligent technology in security, including intrusion detection, IoT security, biometric authentication, and blockchain. He is senior member of IEEE.
\end{IEEEbiography}
 




\vfill

\end{document}